\documentclass[useAMS,usenatbib,usegraphicx]{mn2e}
\usepackage{graphicx}
\usepackage{longtable,lscape}

\def\apss{Ap\&SS}
\def\apj{ApJ}
\def\apjs{ApJS}
\def\mnras{MNRAS}

\title[The sdO pulsator J16007+0748]
      {The rapidly pulsating sdO star, SDSS J160043.6+074802.9}

\author[C. Rodr\'iguez-L\'opez et. al.]
       {C. Rodr\'iguez-L\'opez$^{1,2,6}$, A.E.Lynas-Gray$^3$, D. Kilkenny$^4$, 
J. MacDonald$^5$, A. Moya$^6$, \and C. Koen$^7$, P.A. Woudt$^8$, D.J. Wium$^9$, B. Oruru$^9$ and 
E.Zietsman$^{10}$\\
$^1$Laboratoire d'Astrophysique de Toulouse-Tarbes, Universit\'e de Toulouse, CNRS, Toulouse 31400, France. \\
$^2$Departamento de F\'isica Aplicada, Universidade de Vigo, Vigo 36310, Spain. \\
$^3$Department of Physics, University of Oxford, Keble Road, Oxford OX1 3RH, UK. \\
$^4$Department of Physics, University of the Western Cape, Private Bag X17, Bellville, 7535 Cape, South Africa. \\
$^5$Department of Physics and Astronomy, University of Delaware, Newark, DE 19716, USA. \\
$^6$Departmento de F\'isica Estelar, Instituto de Astrof\'isica de Andaluc\'ia-CSIC, Granada 18008, Spain. \\
$^7$Department of Statistics, University of the Western Cape, Private Bag X17, Bellville, 7535 Cape, South Africa. \\
$^8$Department of Astronomy, University of Cape Town, Private Bag X3, Rondebosch 7701, South Africa. \\
$^9$National Astrophysics and Space Science Programme, Department of Physics, University of the Free State, Bloemfontein 9300, South Africa. \\
$^{10}$Department of Mathematical Sciences, University of South Africa, UNISA 0003, South Africa. \\}

\date{Accepted      
      Received  2009 March 11$^{\rm th}$}

\begin{document}

\maketitle

\begin{abstract}

A spectroscopic analysis of SDSS J160043.6+074802.9, a binary system
containing a pulsating subdwarf-O (sdO) star with a late-type
companion, yields $T_{\rm eff}$ = 70\,000 $\pm$ 5000~K and log $g$ =
5.25 $\pm$ 0.30, together with a most likely type of K3~V for the
secondary star. We compare our results with atmospheric parameters
derived by \citet{fontaine08} and in the context of existing evolution
models for sdO stars. New and more extensive photometry is also
presented which recovers most, but not all, frequencies found in an
earlier paper. It therefore seems probable that some pulsation modes
have variable amplitudes. A non-adiabatic pulsation analysis of
uniform metallicity sdO models show those having $\log g >$ 5.3 to be
more likely to be unstable and capable of driving pulsation in the
observed frequency range.

\end{abstract}


\section{Introduction}

In the course of a search for new AM CVn stars, the Sloan Digital Sky
Survey (SDSS) object, SDSS J160043.6+074802.9, was discovered to be a
very rapid pulsator \citep[][hereafter Paper~I]{woudt06}. Subsequent
observations, reported in the same paper, shows the variability to be
complex, consisting of a strong variation (amplitude $\sim$0.04~mag)
with a period of 119.33~s, its first harmonic (0.005~mag; 59.66~s) and
at least another eight having periods between 62 and 118~s and
amplitudes in the range 0.003 to 0.007~mag.

Preliminary spectroscopy in Paper~I showed the star to be a hot (sdO)
subdwarf with a cooler Main-Sequence companion. The object was known
to be very hot from the SDSS colours ($u'-g' = -0.201$, $g'-r' = -0.205$)
and the spectrograms confirm this, showing He~II 4686\AA~ and the
Pickering series lines of He~II at 5411, 4542 and 4200\AA~ to be
clearly present in absorption. The Balmer series lines H$\beta$ to
H$\epsilon$ are also present, but these are probably blended with
Pickering lines of He~II. No He~I lines (such as the typically strong
features near 4471 and 4026\AA) are detected, ruling out types sdB and
sdOB. The relatively narrow absorption lines also seem to rule out the
possibility of the object being a white dwarf star. In the red region,
the Na~D lines (5889 and 5896\AA) are present and -- given the
presence of He~II lines -- indicate that the object is a binary. This
was confirmed by changes in radial velocities determined from two
spectrograms obtained a few weeks apart, though the data were
insufficient to even suggest an orbital period.  The spectrum is
apparently that of a classical sdO star \citep[see, for
example,][]{moehler90} combined with a late-type Main-Sequence star.

For convenience, we abbreviate the object name, SDSS
J160043.6+074802.9, to J16007+0748. The 2000 equatorial co-ordinates
are implicit in the full SDSS name; the equivalent galactic
co-ordinates are $\ell = 18.65^{\circ}; b = +41.27^{\circ}$. The Sloan
Digital Sky Survey gives photometry: $u' = 17.21, g' = 17.41, r' = 17.61,
i' = 17.66, z' = 17.77$; for more details and background, see Paper~I.

In this paper we present a spectroscopic determination of effective
temperature $(T_{\rm eff})$, surface gravity $(\log g)$ and helium
abundance $(N(He)/N(H))$ using a flux calibrated SDSS spectrum. We
also present new and more extensive photometry for J16007+0748, which
recovers most, but not all frequencies found in Paper~I; we conclude
that some of the pulsation modes have variable amplitudes. Several
uniform metallicity sdO star models, having $T_{\rm eff}$ and $\log g$
within the 68 per cent confidence limits derived for these quantities,
were tested for non-adiabatic pulsation stability. However, pulsations
are not truly driven in our models, in contrast with
\citet{fontaine08}  non-uniform metallicity sdO models, which achieve
real destabilisation.

\section{Spectroscopic Analysis}

In this section, we describe an analysis of a SDSS spectrogram of the
system taken from the sixth data release of the Sloan Digital Sky
Survey \citep{adelmann08}.

A late-type companion (Paper~I) with a luminosity high enough to
produce observable Na~I~D lines, and insufficient spectra with which
to determine an orbital period, meant that the procedure of
\citet{simon94} could not be used to disentangle contributions made by
the two stars at each wavelength.  Because a flux calibrated spectrum
of J16007+0748 had been obtained on 2006 April 28th in excellent
weather, as part of the SDSS survey, the \citet{odonoghue97} technique
for analysing energy distributions of composite sdB stars was
developed and applied.  Specifically, as Paper~I identifies
J16007+0748 as a sdO and late-type Main Sequence binary, the
\citet{odonoghue97} method was extended to simultaneously extract all
parameters of interest from the SDSS energy distribution; these are
the interstellar reddening, $T_{\rm eff}$,
angular radii $(\theta)$ and radial velocities $(v)$ of both stars, as
well as the $\log g$ and helium abundance
$(N(He)/N(H))$ of the sdO star.

The program {\scriptsize TLUSTY} \citep{hubeny95} was used to compute
a grid of model stellar atmospheres for the sdO in Non-Local
Thermodynamic Equilibrium (non-LTE), comprising hydrogen and helium
only; this was justified only in so far as lines attributable to other
elements, and originating in the sdO star spectrum, are not identified
in Paper~I spectra.  Standard model atoms for H~I, He~I and He~II were
adopted with the number of non-LTE levels being 9, 14 and 14
respectively. Level populations for all non-LTE levels were calculated
with  occupation probabilities taken into account following
\citet{hubeny94}; this also applies to the calculation of synthetic
spectra described below.

With computed model atmospheres, non-LTE synthetic spectra (3500
$\leqslant$ $\lambda$ $\leqslant$ 6800{\AA} and having hydrogen and
helium lines only) were calculated using the program {\scriptsize
  SYNSPEC} by Hubeny \& Lanz \citep[see][and references
  therein]{zboril96} with revisions to He I line broadening as
described by \citet{mortimore06}.  Non-LTE synthetic hydrogen and
helium line spectra were produced for the sdO star for the wavelength
range $3500 \leqslant \lambda \leqslant 6800${\AA}.  Balmer line
broadening was based on the \citet*{vidal70} static ion approximation
as \citet{schoning94} shows ion dynamics mainly affect line centres
and are masked in stellar atmospheres.  He~II line broadening was
calculated using tables by \citet{schoning89}.  For comparison with
observation, theoretical spectra were convolved with the appropriate
wavelength dependent instrumental profile; for reasons discussed
below, the projected stellar rotation velocity $(v\sin i)$ was taken
to be zero. Interstellar reddening was modelled following
\citet{seaton79} and \citet{howarth83}.

An initial manual comparison of observed and theoretical spectra was
based on the grid $55\,000 \leqslant T_{\rm eff} \leqslant 70\,000$~K
$({\Delta}T_{\rm eff} = 5000)$, $5.0 \leqslant \log g \leqslant 6.5$
$({\Delta}\log g = 0.5)$ and $0.1 \leqslant N(He)/N(H) \leqslant 0.5$
$({\Delta}(N(He)/N(H))=0.2)$; from this it became clear the sdO star
had a $T_{\rm eff}$ near or above the upper limit of the initial grid.
After some trials, the final search was based on linear interpolation
in the parameter space cube having $(T_{\rm eff}, \log g, N(He)/N(H))$
coordinate extremes specified by (66\,000~K,4.9,0.2) and (76\,000~K,6.5,0.5).

For the late-type companion, synthetic spectra by \citet{martins05}
were used, in preference to atlases derived from observed spectra,
because there is then a well defined $T_{\rm eff}$ scale on which to
base interpolation.  Paper~I results suggest the companion is a late F
or early G-type Main Sequence star.  Accordingly, synthetic spectra
for the late-type companion were generated by piecewise linear
interpolation in the range $3000 \leqslant T_{\rm eff} \leqslant
9500$~K; $\log g = 4.5$ and solar metallicity  being adopted in the
first instance.  Again the projected stellar rotation velocity was
assumed to be zero and synthetic spectra were convolved with the
appropriate wavelength dependent instrumental profile.

Trial radial velocities between $-200$ and $+200$~km/s were selected
for both components of the binary.  The maximum flux density among all
sdO star model spectra was identified, and divided by the maximum flux
density in the observed energy distribution; this ratio when
multiplied by 10.0 corresponded to the maximum trial angular radius for
the sdO star.  A maximum trial angular radius for the late-type
companion was similarly selected, except that the almost arbitrary
scaling factor chosen in this case was 50.0 to reflect the smaller
contribution to the observed energy distribution and the longer
wavelength of the peak flux density in its energy distribution.  Zero
was selected as the minimum trial angular radius for both stars and
trial interstellar reddening values were in the range $0.0 \leqslant
E_{B-V} \leqslant 0.4$.  Because interstellar reddening may be
significant, and associated interstellar Ca~II and Na~I D lines
present in the data, the Ca~II (K) line and central pixels of both
Na~I D lines were ignored during the fitting process.

Shifting and scaling appropriately chosen theoretical spectra by
correct radial velocities and squared angular radii respectively, then
adding them together and adjusting for interstellar reddening, should
reproduce the observed energy distribution; this fitting process was
carried out using a revision (Version 1.2) of the genetic algorithm
which \citet{charbonneau95} codes as {\scriptsize PIKAIA}.  Briefly,
nine unknown parameters were regarded as the ``genetic signature'' of
an individual whose fitness was determined by how well the observed
energy distribution was reproduced; this was characterised by the
inverse chi-squared $(\chi^{-2})$.  The principle of natural selection
means fitter individuals are more likely to breed and produce fitter
offspring.  Continuing evolution of a population of individuals over
many generations eventually locates a global minimum $\chi^2$ which
represents the best obtainable fit.

On the basis of constraints discussed above, a random number generator
was used to create populations of individuals.  Initially populations
of 100 individuals, selected with 100 different random number
generator seeds, would be evolved for a 100 generations; those that
evolved most rapidly would be selected for further runs, evolving them
over larger numbers of generations.  The global minimum was identified
as the almost unique $\chi^2$ reached after evolving three distinct
populations for 30\,000 generations. Further tests, varying metallicity
and surface gravity of the late-type companion gave less satisfactory
fits. Fig.~1 shows the observed SDSS energy distribution and deduced
contributions to it from both binary components.

\begin{figure}
\begin{center}
\includegraphics[width=70mm,angle=-90]{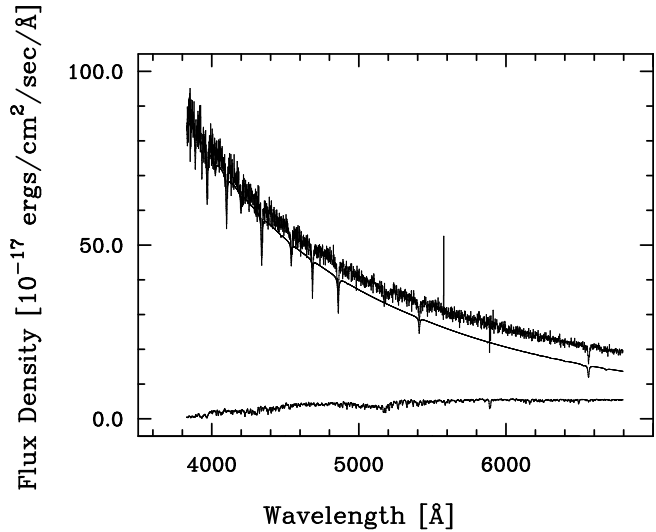}
\caption{Observed SDSS energy distribution and estimated contributions
from the binary components.}
\end{center}
\end{figure}

Standard deviations were estimated following \citet{avni76}, having
regard for the simultaneous estimation of nine parameters.
Hypersurfaces of equal $\chi^2$ in nine-dimensional parameter space
were constructed from all trial solutions generated during an
evolution of 30\,000 generations.  For nine degrees of freedom, one
standard deviation corresponds to $\Delta(\chi^2) = 10.5$.  Parameter
values and standard deviations are presented in Table~1, where it
should be noted that heliocentric corrections have not been applied to
the derived radial velocities.

In Fig.~2, a thin line shows the observed SDSS energy distribution
compared with the sum of the sdO and late-type Main Sequence energy
distributions (shown in Fig.~1) plotted as a thicker line; the
``spike'' near 5600{\AA} has been removed.  Note that the strong Ca~II
K-line near 3933 {\AA} is not fitted and H$\epsilon$ is predicted to
be weaker than actually observed.  There is clearly an interstellar
contribution but also note that Ca~II (K) appears anomalously strong
in the SDSS spectrum when compared with the Paper~I (fig.~6) spectrum.

\begin{figure}
\begin{center}
\includegraphics[width=80mm]{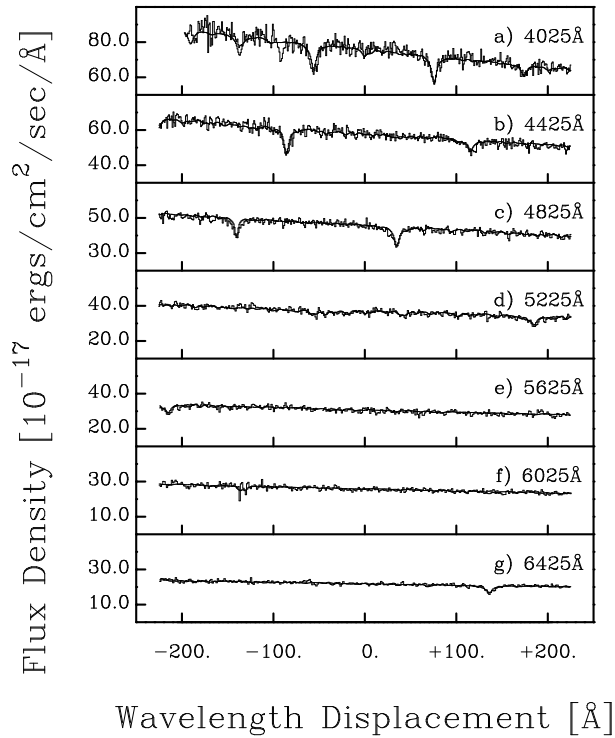}
\caption{Fits to energy distribution of J16007+0748.
Central wavelengths, corresponding to a displacement of 0.0{\AA}
are entered in each panel.}
\end{center}
\end{figure}

\begin{table}
\caption{Derived Parameter Values and Standard Deviations}
\begin{tabular}{l}
\hline
$T_{\rm eff}({\rm sdO}) = 70\,000 \pm 5000$~K \\
${\log}g({\rm sdO}) = 5.25 \pm 0.30$ \\
$N(He)/N(H)({\rm sdO}) = 0.31 \pm 0.11$ \\
$\theta({\rm sdO}) = 6.27 \pm 0.25 \times 10^{-13}$ radians \\
$v({\rm sdO}) = -66 \pm 26 $~km/s \\
$T_{\rm eff}({\rm K3 V}) = 4690 \pm 75$~K \\
$\theta({\rm K3 V}) = 5.26 \pm 0.15 \times 10^{-12}$ radians \\
$v({\rm K3 V}) = -67 \pm 32 $~km/s \\
$E_{B-V} = 0.173 \pm 0.013$ \\
\hline
\end{tabular}
\end{table}

Otherwise the fit shown in Fig.~2 appears to be satisfactory. In
particular, features near 5100 {\AA} attributable to the late-type
companion are reproduced and justify $T_{\rm eff}({\rm K3~V})$ in
Table~1 rather than the higher value (corresponding to a late F or
early G Main Sequence star) which Paper~I suggests.  Moreover there
appears to be little prospect of obtaining a superior fit to the SDSS
energy distribution using theoretical profiles additionally broadened
as a consequence of stellar rotation; for this reason, the assumption
of $v \sin i = 0.0$ for both stars was appropriate.

For 2326 degrees of freedom (the number of wavelength points used less
the number of free parameters), $\sqrt(2{\chi}^2) = 67.88$ was
obtained; statistically $\sqrt(2{\chi}^2)$ should approximate to a
Gaussian distribution having a mean of 68.21 and  unit
variance. Counts in adjacents spectrum pixels are correlated and
so a good fit may not be verified using the $\chi^2$-test.  With
the Runs Test \citep{wald40} the expected number $(<r>)$ of
alternating sequences of positive and negative fit residuals was
estimated.  For the Fig.~2 fit, with 2335 valid data points,
$<r>=1166.5 \pm 24.1$ (s.d.);  the observed number of runs was
$r=914$ which is significantly smaller than $<r>$, as may be
expected (in part at least) having used H/He synthetic spectra to
represent the sdO star which would be expected to have weak
(unobserved) metal lines.

\section{Interstellar Redenning}

In order to check the Section~2 solution, and determination of
$E_{B-V}$ in particular, populations of 100 individuals were evolved
for 10\,000 generations having $E_{B-V}$ fixed at 0.0, 0.044 and
0.173.  $E_{B-V}=0.044$ was chosen because it is predicted by the
adopted galactic extinction model as explained below. All other
parameters were free, with the exception of $T_{\rm eff}$, which was
fixed at 70\,000~K for reasons discussed below, with values determined
by attempting to fit the observed energy distribution as already
described.  For $E_{B-V} = 0.0$, 0.044 and 0.173 the values of
$\sqrt(2{\chi}^2)$ obtained were 74.87, 71.81 and 67.97 respectively;
these are 6.66, 3.60 and -0.23 standard deviations away from the
expected 68.21.  As already noted, individual fluxes in adjacent
pixels of the SDSS spectrum are not independent and so $E_{B-V} = 0.0$
or 0.044 cannot be ruled out with a high degree of confidence;
although it was clear a satisfactory fit could not be obtained with
these interstellar extinctions.

The observed energy distribution is strongly dependent on $E_{B-V}$
and, for a hot star such as J16007+0748, only weakly dependent on
$T_{\rm eff}$.  $T_{\rm eff}$(sdO) and $\log g$(sdO) are obtained from
simultaneously fitting available Balmer lines.  Forcing $E_{B-V}=0.0$
or $0.044$, with $T_{\rm eff}$(sdO) unconstrained, inevitably drove
solutions towards regions of parameter space where $T_{\rm eff}$(sdO)
was unphysically low, given the observed spectrum. Of particular
interest is using $E_{B-V}=0.0$ and $T_{\rm eff}$(sdO)$ = 70\,000$~K,
as it allows a direct comparison with the \citet{fontaine08}
result. We obtained $\log g = 5.37 \pm 0.30$, which is still off by
0.15~dex of their lowest $\log g$ determination.

Interpretation of the derived $E_{B-V}$ requires a knowledge of the
wavelength-dependent light loss (if any) at the spectrograph slit
while the spectrum of J16007+0748 and any flux calibration
standards used by the SDSS project, were being obtained. Bohlin \&
Gilliland's (2004) absolute flux distribution was used with
\citet{smith02} $u'g'r'i'z'$ filter
response curves, and equation 3 in \citet{smith02}, to obtain
synthetic magnitudes for the SDSS standard BD\,$+17^{\rm o}4708$; these
agreed with photometric results within published error limits.
Applying the same approach to the SDSS flux calibrated spectrum of
J16007+0748 gave $(g'-r')=-0.176$ for comparison with the
photometric value of $-0.205$.

Applying the transformation given by \citet{smith02} (their table 7),
gave the corresponding colour correction, due to wavelength-dependent
light loss at the spectrograph slit, of
$\Delta(B-V)=0.030$. Reddening in the direction of J16007+0748 was
therefore estimated to be $E_{B-V}=0.143$.  Of course, the assumption
made here was that J16007+0748 only varies as a consequence of the
pulsation Paper I reports. 

For a high galactic latitude object like J16007+0748, the
axisymmetric model (which does not include a spiral arm contribution)
by \citet{amores05} was appropriate for estimating the galactic
contribution to interstellar reddening along its line of sight; this
gave $E_{B-V}=0.044$ for all distances greater than 500~pc.  The
contribution to $E_{B-V}$ unaccounted for was then 0.099.  Anomalous
reddening in the direction of J16007+0748 is one possibility, as is
the presence of dust and gas in the vicinity of this object which
could have come from the sdO progenitor; these and other possibilities
need a further investigation, which is beyond the scope of the present
paper.

\section{Evolution}

While work reported in this paper was nearing completion,
\citet{fontaine08} published their spectroscopic analysis of
J16007+0748 and explain Paper~I pulsation in terms of radiative
levitation.  The $\log g$(sdO) Fontaine et al. obtain is $5.93\pm0.11$
which disagrees with our Table~1 result, though the corresponding
$T_{\rm eff}$ values are consistent within error limits.  The weakness
of our analysis was the dependence on one SDSS spectrum, and the SDSS
flux calibration in particular.  The strength of the
\citet{fontaine08} work lies in those authors having obtained a high
signal-to-noise spectrum, although of lower resolution than the SDSS
spectrum used in the present work, on which to base their analysis.

\begin{figure}
\includegraphics[width=62mm,angle=-90]{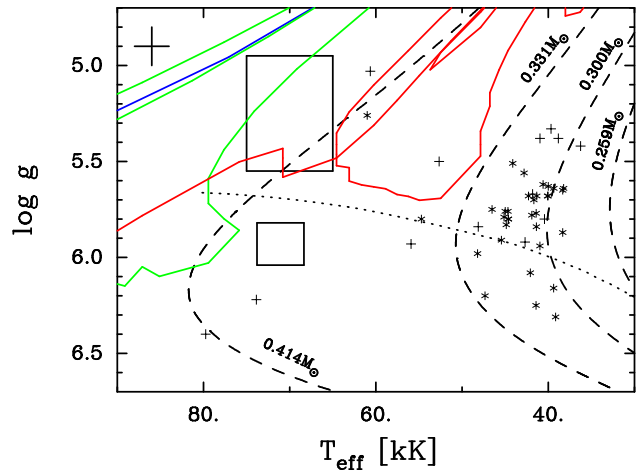}
\caption{Predicted positions of J16007+0748 in a $T_{\rm eff}$--$\log
g$ diagram.  The large upper rectangle shows the position and its
uncertainty as determined in the present paper. The lower and smaller
rectangle is the corresponding position which \citet{fontaine08}
determine. See text  for further details.}
\end{figure}

A sequence of Balmer lines (H$\alpha$ -- H8) provides a $T_{\rm eff}$
and $\log g$ determination as is well known \citep{saffer94}; this
fact allowed authors of the present paper, using the observed energy
distribution, to simultaneously determine $E_{B-V}$ (qualified as
above) which Fontaine {\it et al.}\ implicitly adopt as zero. On the
other hand, although the Fontaine et al.  subtraction of a late-type
template spectrum is not quite perfect (notice the residual Na~I~D
line in the sdO star spectrum in their fig.~1), they do identify the
late-type companion as a G0~V star which is consistent with Paper~I.
Given the similar $T_{\rm eff}$(sdO) determinations obtained by
\citet{fontaine08} and ourselves, the accounting for interstellar
reddening in our study required a redder model of the observed energy
distribution; this was achieved through a redder late-type companion,
which might qualitatively account for the difference between the two
analyses.

Determinations of $T_{\rm eff}$ and $\log g$ for J16007+0748 are
compared with those for 46 sdO stars by \citet{stroeer07} in Fig.~3.
Helium-deficient sdO stars are plotted as ``$+$'' and helium-enriched
sdO stars as``$*$''.  Uncertainties in \citet{stroeer07} measurements
are represented by a cross plotted in the top left-hand corner.  Also
shown in Fig.~3 as solid lines are three Post Asymptotic Giant Branch
(AGB) tracks by \citet{blocker95a, blocker95b}; different tracks are
distinguished by different colours (blue - hydrogen burning
$0.605~M_{\odot}$, green - helium burning $0.625~M_{\odot}$ and red -
helium burning $0.524~M_{\odot}$). Dashed lines are used to plot
helium white dwarf evolution tracks by \citet{driebe98,driebe99} and
the position of the Zero-Age Helium Main Sequence shown with a dotted
line.  Each evolution track plotted in Fig.~3 is labelled with the
corresponding stellar mass.

\citet{stroeer07} discuss evolutionary scenarios for both
helium-deficient and helium-enriched sdO stars.  J16007+0748 is found
to be helium-enriched by \citet{fontaine08} and confirmed, in
principle at least, by results in Table~1.  Only Stroeer et
al. scenarios concerned with helium-enriched sdO evolution were
therefore considered.

\citet{driebe98} evolve $1~M_{\odot}$ stars from pre-Main Sequence to
the Red Giant Branch (RGB); their envelopes are almost completely
stripped by Roche-lobe overflow in a close binary system before core
helium burning starts.  The remaining hydrogen-shell turns out to be
rather thick and it is not obvious how a helium-enriched atmosphere is
obtained.  If the residual mass lies in the range $0.21~M_{\odot}
\leqslant M \leqslant 0.3~M_{\odot}$, \citet{driebe98} show that
hydrogen-shell flashes occur which may lead to a dredge-up of helium,
but predicted positions of J16007+0748 in Fig.~3 are inconsistent with
Driebe et al. tracks having these masses.  If J16007+0748 is a
post-RGB object, it has to have a mass  $\sim 0.4~M_{\odot}$ and it is
not then clear how helium-enrichment could have occurred.  Continued
stripping of the hydrogen envelope by interaction with the late-type
companion is a possibility though this would have speeded up
evolution, because most of the luminosity comes from the hydrogen
burning shell, making detectability less likely.

Positions of J16007+0748 in Fig.~3 are also close to post-AGB tracks
(not shown) by \citet{schonberner79,schonberner83} and, like the
helium-enriched sdO HE\,1430--0815 which \citet{stroeer07} consider,
it could be a post-AGB star, but the timescale for evolution from the
AGB to the pre-white dwarf stage \citep{schonberner79} is only $\sim 3
\times 10^4$ years and again this reduces the probability of
detection.  Unfortunately the pulsation mass of $0.48~M_{\odot}$ by
\citet{fontaine08}, which these authors choose to be consistent with
typical sdB masses and therefore needs improvement as they
acknowledge, did not enable us to distinguish between the post-RGB
$(M\sim0.4~M_{\odot})$ and post-AGB scenarios $(M\sim0.55~M_{\odot})$.

\citet{blocker95a, blocker95b} develops the earlier work by
\citet{schonberner79,schonberner83} by considering evolution of low
and intermediate mass stars with mass-loss taken into account with
different descriptions for the RGB, AGB and post-AGB phases. Three of
Bl{\"o}cker's post-AGB tracks pass close the error rectangles in
Fig.~3.  The hydrogen burning 0.605~$M_{\odot}$ track (shown in blue),
for which the progenitor mass was 3~$M_{\odot}$, passes near the top
left-hand corner of the error rectangle obtained in the present paper,
which suggested the sdO in J16007+0748 would have to have $M \leqslant
0.605 M_{\odot}$ if it is a hydrogen burning post-AGB star.  The
post-AGB age would be only $\sim 10^4$ years and detecting such an
object has to be considered unlikely.

Of greater interest are the 0.524~$M_{\odot}$ and 0.625~$M_{\odot}$
helium burning models which have three branches in Fig.~3 (plotted in
red and green respectively). Helium burning arises because the last
thermal pulse occurs in the post-AGB phase, transforming the hydrogen
burner into a helium burner; the consequence is rapid evolution back
to the AGB where post-AGB evolution begins once more as a helium
burner.  The longer post-AGB evolution time implied by helium burning,
through crossing the $T_{\rm eff} - \log g$ space defined by Fig.~3
three times, makes detection more likely.  J16007+0748 could therefore
have a sdO star which is a post-AGB helium burning object with $M \sim
0.6~M_{\odot}$; the luminosity would the be $\sim 10^3~L_{\odot}$
which was consistent with observation as discussed below.

Radial velocities obtained for the K3~V and sdO stars were found to be
similar, although with large standard deviations, suggesting they
constitute a physical binary.  The absolute magnitude $(M_V)$ of the
K3~V star indicated a distance $(d)$ of $\sim 3$~kpc.  Angular radii
then gave radii of $\sim 0.08$ and  $\sim 0.7~R_{\odot}$ for the sdO
and K3~V stars respectively, with a corresponding sdO star luminosity
$\sim 150~L_{\odot}$.  The Table~1 $\log g$ (sdO) resulted in a sdO star
mass of  $\sim 0.04~M_{\odot}$.  For the sdO star to have a mass
nearer $\sim 0.4~M_{\odot}$, for reasons discussed above, its distance
needs to be about a factor of three larger.

A possible association of the sdO star with post-RGB \citep{driebe98}
or post-AGB (\citealt{blocker95a,blocker95b}; \citealt{schonberner79,
  schonberner83}) evolutionary tracks would imply respective
luminosities of $\sim 10^{2.9}$ and $\sim 10^{3.1}~L_{\odot}$,
consistent with $d \sim 9$~kpc. A $d \sim 9$~kpc places J16007+0748,
given its galactic coordinates at $\sim 7$~kpc from the Galactic
Centre.  For J16007+0748 to be a physical binary, the late-type
companion would then have $M_V \sim 4.2$ rather than the expected $M_V
\sim 6.7$ for a normal K3~V star.

A higher $M_V$ for the late-type companion could be explained if the
Paper~I and \citet{fontaine08} classifications for the late-type
companion are correct, and not the one presented in this paper; for
this to be satisfactory, corroboration is needed for the implicit
Fontaine et al.  assumption of $E_{B-V} = 0.0$. If the late-type
companion had evolved off the Main Sequence and is in reality a K3~IV
star, the higher luminosity and radius might be explained; but this is
also unsatisfactory because the Section 2 analysis found less
satisfactory fits to the SDSS energy distribution were obtained when
lower gravity synthetic spectra were used for the late-type companion.
Accretion of material from the sdO progenitor could result in the K3~V
star appearing to be over-luminous; a theoretical justification would
be needed and is beyond the scope of the present paper.

\section{New photometry}

New photometry was obtained with the same instrumentation and methods
used for the observations presented in Paper~I. In brief, observations
were made during a ``core'' run of fourteen consecutive nights in 2006
June/July, followed by three consecutive nights one week later and a
further two nights about a month later. All observations were made with
the University of Cape Town CCD photometer
\citep[UCTCCD;][]{odonoghue95} on the 1.9-m telescope at the
Sutherland site of the South African Astronomical Observatory (SAAO).
The UCTCCD was used in frame-transfer mode which has the advantage of
losing no time on target when the CCD reads out, enabling continuous
or ``high-speed'' observing.

\begin{table}
\centering
\caption{Observation log for J16007+0748 (1.9-m telescope + UCTCCD).}
\begin{tabular}{|rcccl|}
\hline
     Date   &   JD      &  Run   & Comments  \\
     2006   &           &  (hr)  &           \\
\hline
Jun 28/29 &  245~3915 &  5.6   &               \\
    29/30 & ~~~~~3916 &  5.7   &               \\
    30/01 & ~~~~~3917 &  5.7   &               \\
Jul~~~~~1/2& ~~~~~3918&  5.5   &               \\
     2/3  & ~~~~~3919 &  5.5   & software crash\\
     3/4  & ~~~~~3920 &  5.6   &               \\
     4/5  & ~~~~~3921 &  1.7   & cirrus        \\
     5/6  & ~~~~~3922 &  2.8   & cirrus        \\  
     7/8  & ~~~~~3924 &  5.1   &               \\
     8/9  & ~~~~~3925 &  3.4   & some cloud    \\
     9/10 & ~~~~~3926 &  4.9   & cirrus        \\
    10/11 & ~~~~~3927 &  5.2   &               \\
    11/12 & ~~~~~3928 &  5.2   &               \\
          &           &        &               \\
    16/17 & ~~~~~3933 &  2.8   &               \\
    17/18 & ~~~~~3934 &  5.0   &               \\
    18/19 & ~~~~~3935 &  4.7   &               \\
          &           &        &               \\
Aug 17/18 & ~~~~~3965 &  2.1   &               \\
    20/21 & ~~~~~3968 &  2.2   &               \\
\hline
\end{tabular}
\end{table}

On the 1.9-m telescope, the 22~$\mu$ pixels of the CCD are equivalent
to 0.13 arcseconds at the detector, so that it is normal to use at
least 3 x 3 prebinning for optimal data extraction, unless the seeing
is better than about 1 arcsecond. The first week was almost uniformly
good and all observations were made with 3 x 3 pre-binning; the second
week was less good, though mostly usable, and most of the observations
were made with  4 x 4 pre-binning. All integration times were 10
seconds, except for the  runs on June 28/29 and July 4/5, where a time
of 12 seconds was used. These  values were considered a reasonable
compromise between obtaining as good a signal-to-noise as possible and
having adequate temporal sampling to resolve rapid
variations. Conventional procedures (bias subtraction, flat field
correction and so on) were followed with magnitude extraction being
based on the point-spread function of the {\scriptsize DoPHOT} program
as described by \citet*{schechter93}.

All photometry of J16007+0748 was corrected differentially to remove
any rapid transparency variations using the nearby bright star (about
20 arcseconds SE of J16007+0748; see the chart given in fig.~1 of
Paper~I).  Since, in general, field stars will be quite red, we might
expect differential extinction effects to be significant, and we have
further corrected for residual atmospheric extinction effect by
removing a second order polynomial from each night's
observations. This means that we might be removing real changes in
stellar brightness on time scales of a few hours but this has to be
accepted.

\begin{figure}
\begin{center}
\includegraphics[width=80mm]{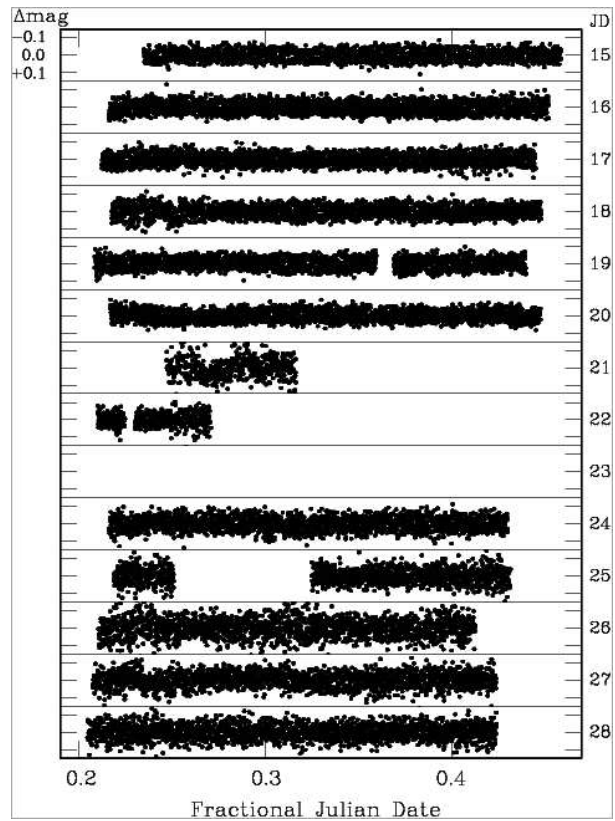}
\caption{2006 June/July observations for J16007+0748. Numbers at the
right-hand edge are JD -- 2453900.}
\end{center}
\end{figure}

A log of the ``high-speed'' or continuous photometry obtained for
J16007+0748 is given in Table~2. (Note that runs much longer than about 6
hours are impossible for a star at declination $\sim +8^{\circ}$ from
Sutherland without reaching undesirable air mass values). A sample of a
differentially corrected light curve is shown in fig.~2 of Paper~I; that
light curve is very similar to all the photometry analysed here.

Fig.~4 shows the distribution in time of the 2006 June/July observations. 
It is clear that they fall into two main sections, the first six nights 
and the last five. It can also be seen from the figure that the scatter
is rather worse in the last five nights -- a result of somewhat poorer
weather and seeing conditions, on average.

\section{Frequency analysis}

The frequency analyses described in this section were carried out using
software which produces Fourier amplitude spectra following the Fourier 
transform method of \citet{deeming75} as modified by \citet{kurtz85}.

\subsection{The ``core'' data}

The observations shown in Fig.~4 lend themselves in an obvious way to our
preferred practice of reducing the observations initially in two approximate
``halves''. We have taken the first six nights and the last five nights and
determined the Fourier amplitude spectra for these separately. The baselines
of the two data sets correspond to resolutions of about 2.2 and 2.7~$\mu$Hz
respectively, and the rms scatter in the background noise of the Fourier
amplitude spectra (avoiding any obvious signal spikes) are about 0.4 and
0.7~mmag. Fig.~5 shows the results after pre-whitening each set by the
dominant frequency near 8380~$\mu$Hz to make the weaker frequencies more
visible. The difference in background noise is also clear in that figure
and is a result of the poorer quality of the second half of the observations
(See Fig.~4).

We have initially extracted frequencies from the amplitude spectra one at a
time and then -- for several frequencies common to both sets of observations
-- have carried out simultaneous least-squares fits \citep{deeming68} to the
extracted frequencies, followed by a search for any remaining common
frequencies. By continuing this procedure we find six well-established
frequencies which are listed in Table~3.

The Fourier amplitude spectra of the observations with six frequencies
removed are shown in Fig.~6, in which it appears that there might still
remain common frequencies. Such were not immediately obvious in our initial
search but, allowing for 1 or 2 cycle/day aliasing (corresponding to
differences of 11.6 and 23.1~$\mu$Hz), we can find three more common
frequencies which are also listed in Table~3. The final column of that
table lists the frequency designations, $f_i$, given in Paper~I. 

Using Paper~I nomenclature, we easily recover $f_1$ to $f_5$; $f_6$ --
9651.9~$\mu$Hz in Paper~I -- is almost certainly the peak we find at
9663.0~$\mu$Hz with a one cycle d$^{-1}$ alias either in Paper~I or
here. The peak near 13074~$\mu$Hz is probably $f_7$ with a two cycle
d$^{-1}$ alias, but could also be $f_{13}$ (13072~$\mu$Hz in
Paper~I). Somewhat surprisingly, $f_8$ is a marginal detection in the
current work and we do not recover $f_9$ -- $f_{12}$ even at only
twice the background noise in the present data, whereas they were
detected in Paper~I at four to six times the background. Curiously,
$f_{14}$ does seem to be present and at about the same amplitude as in
Paper~I.

In a sense, this is a disappointing result -- with far more observations
over a slightly longer baseline than the Paper~I data, we had expected to
recover all the original frequencies and hoped to find even more. The
current result probably reflects the fact that some of the frequencies have
variable amplitude; this was evident already in Paper~I (see fig.~3 of that
paper) and is also suggested in our Fig.~5 and Table~3 where the peak near
13074~$\mu$Hz doubles in amplitude between the first six nights and the last
five; in a similar comparison, the other peaks have constant amplitude
within about 2~$\sigma$ to 3~$\sigma$.

\begin{figure}
\begin{center}
\includegraphics[width=80mm]{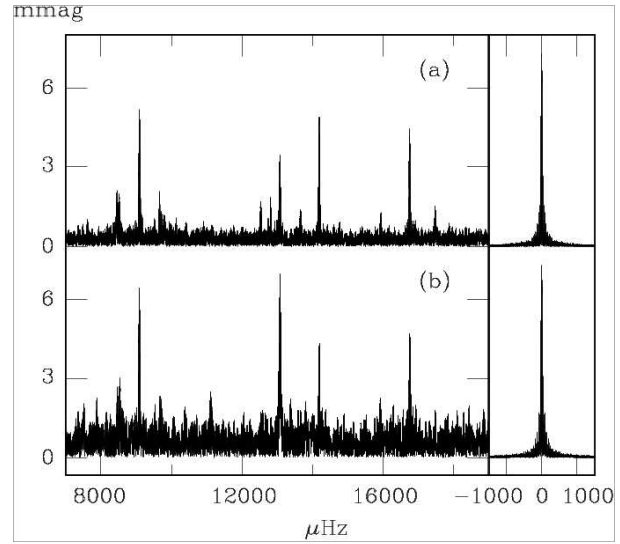}
\caption{Fourier amplitude spectra for (a) the first six nights and (b)
the last five nights (see Fig.~4). Both have been pre-whitened by the
very strong frequency near 8380~$\mu$Hz, better to display the weaker 
frequencies.The corresponding spectral windows are plotted at the right
with the same frequency scale and normalised amplitudes.}
\end{center}
\end{figure}

\begin{figure}
\begin{center}
\includegraphics[width=80mm]{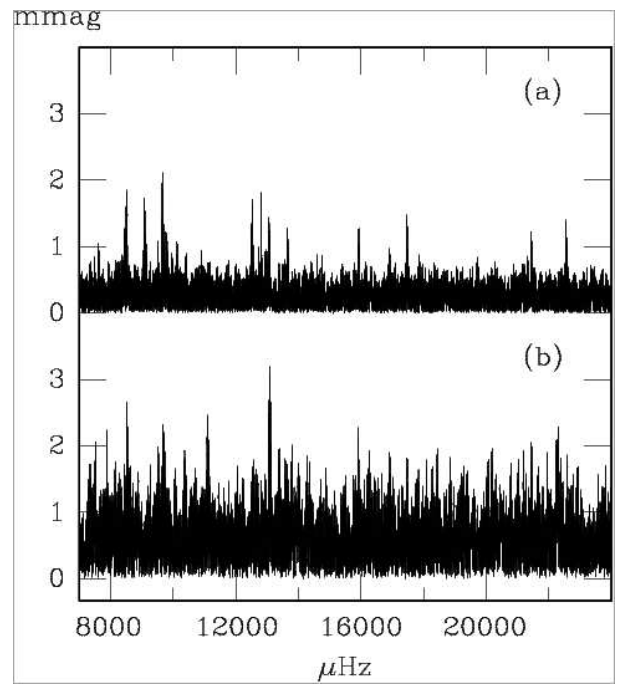}
\caption{Fourier amplitude spectra for (a) the first six nights and (b)
the last five nights (see Fig.~4) after the removal of six common 
frequencies.}
\end{center}
\end{figure}

\begin{table}
\centering
\caption{Six frequencies and amplitudes extracted simultaneously plus
three frequencies extracted singly (and allowing for aliasing) from the
analysis represented in Figures 5 and 6. Formal errors in the amplitudes are
$\sim$ 0.3~mmag for the first 6 nights and $\sim$ 0.6~mmag for the last five
nights.}
\begin{tabular}{|ccccccc|}
\hline
\multicolumn {3}{|c|}{First 6 nights} &
\multicolumn {3}{|c|}{Last 5 nights} \\
 Freq    & $\pm$   & Amp  &  Freq    & $\pm$   & Amp   &  Paper I \\
$\mu$Hz  & $\mu$Hz & mmag & $\mu$Hz  & $\mu$Hz & mmag  &          \\
\hline
~8379.48 & 0.01 & 38.1 & ~8379.53 & 0.02 &  38.0 &  $f_1$        \\
~9091.32 & 0.05 & ~5.2 & ~9091.08 & 0.11 &  ~6.5 &  $f_2$        \\
16758.91 & 0.06 & ~4.2 & 16759.00 & 0.15 &  ~4.7 &  $f_3$        \\
14186.28 & 0.05 & ~4.9 & 14186.86 & 0.16 &  ~4.3 &  $f_4$        \\
13073.82 & 0.08 & ~3.4 & 13074.67 & 0.10 &  ~7.0 &  $f_7/f_{13}$ \\
~8461.7~ & 0.12 & ~2.2 & ~8462.4~ & 0.25 &  ~2.8 &  $f_5$        \\
         &      &      &          &      &       &               \\
~9663.0~ & 0.14 & ~2.1 & ~9686.2~ & 0.34 &  ~2.4 &  $f_6$ ?      \\
~8506.3~ & 0.14 & ~1.9 & ~8510.5~ & 0.33 &  ~2.1 &  $f_{14}$     \\
15939.3~ & 0.23 & ~1.3 & 15940.4~ & 0.35 &  ~2.3 &  $f_8$        \\
\hline
\end{tabular}
\end{table}

\subsection{The first six ``core'' nights}

Because amplitude variation -- whatever the cause -- will confuse the
analysis, we have taken the first six nights (which are clearly of better
quality than the rest of the observations) and carried out a similar process
to that in Section 6.1, in this case splitting the observations into two
sets of three nights. The resolution for each set is about 5~$\mu$Hz and the
background noise is around 0.5~mmag. This subset of the observations is
comparable with that of Paper~I -- almost 30\% more data but obtained over 
a baseline of six nights rather than ten.

After the removal of eight common frequencies with amplitudes greater than
about four times the background, there still seem to be significant peaks in
the amplitude spectra (Fig.~7). If, as before, we allow for aliasing and also
search down to three times the background (given that we have two separate
sets of data), five more common frequencies are found. Again, we recover
$f_1$ to $f_7$ and $f_{14}$, all at four times the background or better
(see Table~4). In addition, we have weak evidence for $f_8$, $f_{10}$ and
$f_{12}$ and maybe two new frequencies, but these results are not totally
convincing. We still find no significant evidence for the presence of 
$f_9$ and $f_{11}$; this does not mean we doubt their existence in the 
Paper~I data, indeed, they seem well-established there. Rather, it seems 
more likely that we are seeing more evidence for amplitude variation.

\subsection{The June/July and August data}

Analysing separately the three nights from mid-July (16/17 -- 19/20),
we easily recover the frequencies $f_1$ to $f_7$, well above the
background noise (allowing for some one- and two-cycle/day
aliasing). However, when we add these observations to the core data,
the analysis is less clear. The strongest frequencies are still
recovered but appear to be accompanied by weak peaks within about a
$\mu$Hz or so. This immediately suggests that variable amplitudes
might be the cause (whether due to unresolved pairs of frequencies or
actual amplitude variation in a single mode) and that these are made
clearer by the longer baseline. But it is also possible, given that
J16007+0748 is a spectroscopic binary of unknown orbital period (see
Paper~I), that the orbital motion is creating an effective phase shift
in the later results relative to the earlier. In this case, any
derived frequency would be affected and extraction of a sine wave with
a slightly incorrect frequency could result in cyclic residuals in the
data which would then be extracted as a nearby, much weaker
signal. Detailed spectroscopic determination of the orbital period
would help to resolve this issue -- and might even enable the
pulsation phase shifts to be used to measure the size of the projected
orbit by determining the light travel time. This was attempted
by \citet{kilkenny98,kilkenny03} for the eclipsing, pulsating
PG\,1336-018, but resulted only in upper limits of about a light second
for the projected orbital diameter -- but that star has a period of
only about 0.1~d and the problem would be more tractable for a longer
period binary.

The August observations were analysed separately and revealed frequencies
$f_1$ to $f_5$ but both series of observations were short (around 2 hours)
and are separated by a month from the other observations, so we have not
attempted an analysis of the merged data.

\begin{figure}
\begin{center}
\includegraphics[width=80mm]{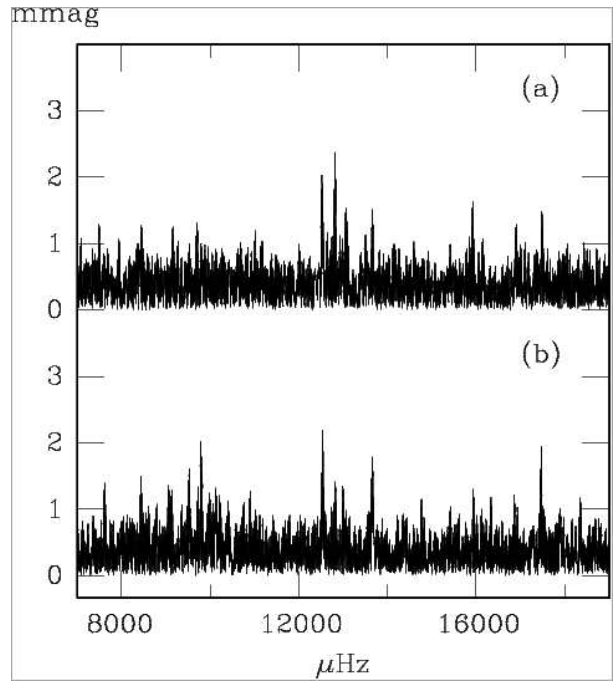}
\caption{Fourier amplitude spectra for (a) the first three nights and (b)
the second three nights (see Fig.~4) after the removal of eight common 
frequencies.}
\end{center}
\end{figure}

\subsection{Summary}

The frequency analysis of the observations reported here has found the
same frequencies $f_1$ to $f_7$ as found in Paper~I. It is possible,
we have also detected $f_8$ and $f_{14}$ and maybe others, though
these detections are much less secure.  It appears, somewhat
paradoxically, that shorter data sets yield ``better'' results and
this is likely to be due to amplitude variation (or equivalently,
frequency beating) or the effects of the binary orbital motion -- and
probably both of those.

\begin{table}
\centering
\caption{Eight frequencies and amplitudes extracted simultaneously from 
the first six nights split into two halves, plus five frequencies extracted
singly (and allowing for aliasing). Formal errors in the amplitudes are
about 0.3~mmag for both halves.}
\begin{tabular}{|ccccccc|}
\hline
\multicolumn {3}{|c|}{First 3 nights} &
\multicolumn {3}{|c|}{Second 3 nights} \\
 Freq    & $\pm$   & Amp  &  Freq    & $\pm$   & Amp   &  Paper I \\
$\mu$Hz  & $\mu$Hz & mmag & $\mu$Hz  & $\mu$Hz & mmag  &          \\
\hline
~8379.48 & 0.02 & 38.3 & ~8379.50 & 0.02 &  38.2 &  $f_1$      \\
~9091.8~ & 0.18 & ~4.1 & ~9090.8~ & 0.12 &  ~6.4 &  $f_2$      \\
16759.0~ & 0.16 & ~4.7 & 16758.9~ & 0.18 &  ~4.1 &  $f_3$      \\
14186.3~ & 0.15 & ~5.0 & 14186.1~ & 0.16 &  ~4.7 &  $f_4$      \\
13074.2~ & 0.16 & ~4.6 & 13072.4~ & 0.29 &  ~2.6 &  $f_7$      \\
~8461.1~ & 0.42 & ~1.8 & ~8460.4~ & 0.24 &  ~3.2 &  $f_5$      \\
~9664.1~ & 0.26 & ~2.9 & ~9662.9~ & 0.40 &  ~1.9 &  $f_6$      \\
~8518.1~ & 0.40 & ~1.9 & ~8516.8~ & 0.36 &  ~2.1 &  $f_{14}$   \\
         &      &      &          &      &       &             \\
12816.5~ & 0.39 & ~2.4 & 12815.2~ & 0.70 &  ~1.3 &             \\
12521.3~ & 0.37 & ~2.0 & 12536.0~ & 0.33 &  ~2.2 &  $f_{12}$ ? \\
17483.9~ & 0.51 & ~1.5 & 17459.0~ & 0.38 &  ~1.9 &             \\
13667.6~ & 0.50 & ~1.5 & 13659.6~ & 0.41 &  ~1.8 &  $f_{10}$ ? \\
15921.6~ & 0.56 & ~1.7 & 15939.7~ & 0.71 &  ~1.3 &  $f_8$      \\
\hline
\end{tabular}
\end{table}

\section{Pulsation modelling}

A pulsation analysis of structural models of
\citet{rod08}\footnote{The evolutionary sequences in this paper
  correspond to \citet{rod08} sequences 10, 11, 12, 13, 14 and 7,
  respectively.  Models 1.1, 3, 4 and 5 in the table correspond,
  respectively, to Models 10.3, 12, 13 and 14 of the same paper.}  was
carried out within the $\log g$--$T_{\rm eff}$ 68 per cent confidence
interval determination for J16007+0748.

\begin{figure}
\includegraphics[width=62mm,angle=90]{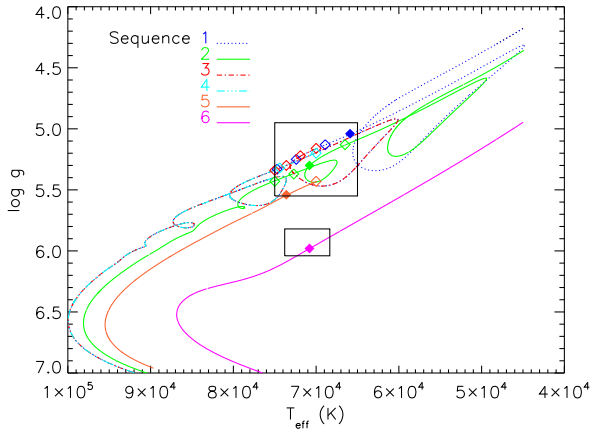}
\caption{Evolutionary tracks and predicted positions of J16007+0748 in
a $T_{\rm eff}$--$\log g$ diagram. Diamonds mark sdO models along the
tracks  analysed in search for instability.   The  analysis of  those
marked with a filled diamond is  explicitly shown in  the paper (see
text  for  further details).   The  large  upper  rectangle shows  the
$\log g$--$T_{\rm eff}$ and its  uncertainty as determined in the
present paper. The lower  and  smaller  rectangle  is the
corresponding  position  which \citet{fontaine08} determine.}
\label{fig:tracks}
\end{figure}

To create the starting points for sequences labeled 1 through 5, a
star of initial mass $1~M_{\odot}$ and initial heavy element abundance
$Z=0.02$ was evolved from the pre-Main Sequence stage to the Red Giant
Branch (RGB), through  the helium core flash and to the Zero Age
Horizontal Branch (ZAHB).  The code used, JMSTAR, is described  by
\citet{lawlor06} and references therein. In order to produce models of
hot subdwarfs \citep{dcruz96}, the mass-loss rate on the  RGB is
enhanced compared to typical observed values, and it is characterised
by a value of the Reimer's mass-loss parameter $\eta_R = 0.675$.   For
comparison, typical observed RGB mass-loss rates correspond to
$\eta_R$ = 0.35 -- 0.4. Because  of the high mass-loss rates, the star
leaves the Giant Branch before the helium core flash occurs. The mass
of the  hydrogen-rich envelope, $\sim 0.01~M_{\odot}$, at this point
is sufficiently small so that the star arrives on the ZAHB relatively
hot, $T_{\rm eff} \sim$ 20\,000~K, and has the characteristics of a
subdwarf-B star. As a greater than solar heavy element abundance in
the driving zone seems necessary for pulsational instability, an ad
hoc extra mixing is introduced for the evolution beyond  the ZAHB. In
the stellar evolution code, convective mixing is modelled by use of
diffusion equations, with the turbulent diffusivity calculated from
mixing length theory. Other mixing processes can be included by adding
terms to the turbulent diffusivity. For our ad hoc mixing, the
turbulent diffusivity is

\begin{figure}
\includegraphics[width=85mm]{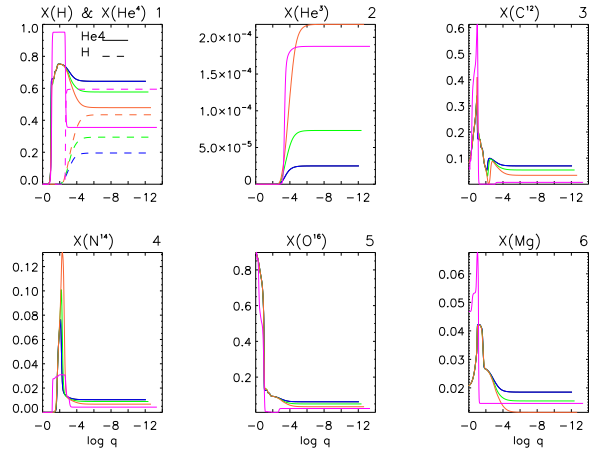}
\caption{Mass fractions  of  the   main  chemical   constituents  of
representative Models: model 1 (blue), 2.1 (green), 5.1 (orange) and 6
(magenta). X(Mg) refers to  the mass fractions of all the other
elements. Top right numbers designate panels' numbers (see text).}
\label{fig:models2}
\end{figure}

\begin{equation}
\sigma _{ad\,hoc} = \frac{1}{\left( 4\pi r^2\rho \right )^2} \frac{\eta M_T}{\tau }
\end{equation}

\begin{figure}
\includegraphics[width=85mm]{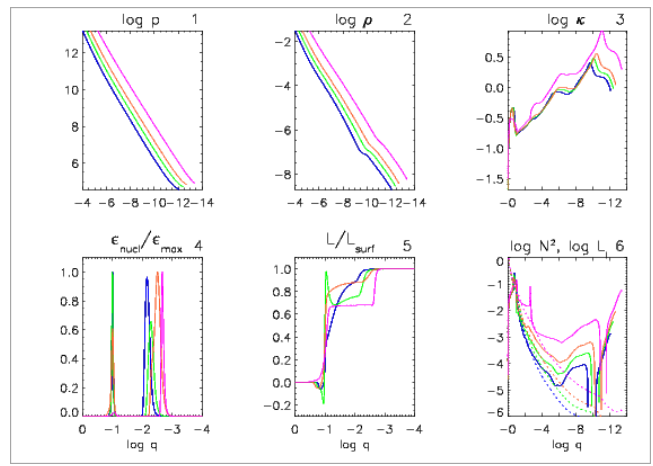}
\caption{Main physical parameters of representative models (colour key
the same as in the other figures). Note that Panels 1 and 2 show only
the outer envelope, while Panels 4 and 5 show only the innermost
regions of the star.}
\label{fig:phys22}
\end{figure}

\begin{table*}
\centering
\caption{Effective temperature, surface gravity and envelope mass  fractions of the
equilibrium  models.  The column  labeled  X(Mg)  refers  to the  mass
fractions of all the other  elements.  Z is the current metallicity of
the model. The capital letter beside some model's number refers to their plotting colours.}
\label{tab:models}
\centering

\begin{tabular}{l|cccccccccc}  
Model  &  T$_{eff}$ & log g  & X(H) & X(He$^3$) &  X(He$^4$) & X(C) &  X(N) & X(O) & X(Mg) & Z \\ 
       &  (K)       &        &      &	      &	             &	    &       &	   &	   &   \\ \hline
1   B  & 65\,900 & 5.04 & 0.20 & 2.5E-05 & 0.64 & 7.0E-02 & 1.0E-02 & 6.2E-02 & 1.9E-02 & 0.16 \\
1.1 & 69\,000 & 5.13 & 0.19 & 2.4E-05 & 0.65 & 7.1E-02 & 1.1E-02 & 6.2E-02 & 1.9E-02 & 0.16 \\ 
1.2 & 72\,400 & 5.25 & 0.20 & 2.2E-05 & 0.65 & 7.2E-02 & 1.1E-02 & 6.3E-02 & 1.9E-02 & 0.15 \\ 
1.3 & 74\,700 & 5.33 & 0.16 & 1.5E-05 & 0.67 & 7.6E-02 & 1.1E-02 & 6.7E-02 & 2.0E-02 & 0.17 \\ 
\hline	
2 & 66\,500 & 5.13 & 0.31 & 8.4E-05 & 0.57 & 5.2E-02 & 8.7E-03 & 4.8E-02 & 1.5E-02 & 0.12 \\ 
2.1 G & 70\,800 & 5.30 & 0.29 & 7.3E-05 & 0.58 & 5.5E-02 & 8.9E-03 & 5.0E-02 & 1.5E-02 & 0.13 \\ 
2.2 & 72\,700 & 5.37 & 0.26 & 5.3E-05 & 0.60 & 6.0E-02 & 9.4E-03 & 5.4E-02 & 1.7E-02 & 0.14 \\ 
2.3 & 75\,000 & 5.43 & 0.26 & 5.0E-05 & 0.60 & 6.1E-02 & 9.5E-03 & 5.4E-02 & 1.7E-02 & 0.14 \\ 
\hline	
3  & 70\,000 & 5.16 & 0.19 & 2.3E-05 & 0.65 & 7.1E-02 & 1.1E-02 & 6.3E-02 & 1.9E-02 & 0.16 \\
3.1 & 71\,900 & 5.22 & 0.19 & 2.2E-05 & 0.65 & 7.2E-02 & 1.1E-02 & 6.3E-02 & 1.9E-02 & 0.16 \\ 
3.2 & 73\,600 & 5.30 & 0.16 & 1.5E-05 & 0.67 & 7.6E-02 & 1.1E-02 & 6.6E-02 & 2.0E-02 & 0.17 \\ 
3.3 & 75\,000 & 5.34 & 0.16 & 1.5E-05 & 0.67 & 7.6E-02 & 1.1E-02 & 6.7E-02 & 2.0E-02 & 0.17 \\ 
\hline	
4  & 70\,000 & 5.20 & 0.16 & 1.6E-05 & 0.66 & 7.6E-02 & 1.1E-02 & 6.6E-02 & 2.0E-02 & 0.18 \\
4.1 & 74\,400 & 5.32 & 0.16 & 1.5E-05 & 0.67 & 7.6E-02 & 1.1E-02 & 6.7E-02 & 2.0E-02 & 0.18 \\ 
\hline	
5  & 70\,000 & 5.43 & 0.48 & 3.0E-04 & 0.45 & 2.8E-02 & 6.0E-03 & 2.9E-02 & 1.0E-02 & 0.07 \\
5.1 O & 73\,600 & 5.54 & 0.43 & 2.2E-04 & 0.48 & 3.4E-02 & 6.7E-03 & 3.4E-02 & 1.1E-02 & 0.09 \\ 
\hline	
6  M & 70\,800 & 5.98 & 0.59 & 1.9E-04 & 0.36 & 7.3E-03 & 4.3E-03 & 2.4E-02 & 1.5E-02 & 0.05 \\ 
\end{tabular}							       
\end{table*}

where $\eta$ is the mass in a zone and $M_T$ the total mass of the
star. The degree of extra mixing is determined by the ratio of the
input time scale $\tau$ to the evolution time scale. A
consequence of extra mixing is that core material is transported
to the surface, increasing heavy element abundances in the
hydrogen-rich envelope. Also, due to mixing of helium to higher
temperatures, the star experiences helium shell flashes  that drive
convective mixing of heavy elements to the surface further increasing
$Z$. The starting points for Sequences 1 and 2  are taken from the
post-ZAHB track when $T_{\rm eff}$ = 45\,000 K. Similarly the starting
points for Sequences 3, 4 and 5 are taken from the post-ZAHB track
when $T_{\rm eff}$ = 70\,000 K. Extra mixing is turned off for the
evolution from these starting points.  For Sequences 2 and 5 the
mass-loss rate is set to zero. For Sequences 1, 3 and 4 mass-loss from
a line-driven wind is assumed with  the rate taken from
\citet{abbott82} but scaled in proportion to $Z^{0.5}$. All models at
this point of evolution have  mass $0.479~M_{\odot}$.

The starting point for Sequence 6 is from the track for a star of
initial mass $1~M_{\odot}$, initial metallicity $Z=0.05$, and $\eta_R
= 0.6$.  Since the heavy element abundance is greater than solar, no
extra mixing was used. The evolution of this model star is more normal
in that it is the helium core flash that terminates the RGB. At the
ZAHB, the hydrogen envelope has a mass  $\sim 0.06~M_{\odot}$ and hence
the star is cool with $T_{\rm eff} \sim$ 4300 K. However, due to
mass-loss primarily during the helium shell burning phase, the mass
of the hydrogen envelope is sufficiently reduced to avoid the AGB
phase. The starting point from Sequence 6 is taken at $T_{\rm eff}$  =
45\,000 K as the star is evolving at roughly constant luminosity to
the blue, on its way to becoming a white dwarf. At this point  the
stellar mass is $0.497~M_{\odot}$.

The evolutionary tracks described by each evolutionary sequence in the
$\log g$--$T_{\rm eff}$ diagram are shown in Fig.~\ref{fig:tracks}.
We note that Sequences 1, 3 and 4, describe the same evolutionary
track, although equivalent models have different envelope
compositions.  A stability analysis was carried out for models marked with diamonds. 
Below, the analysis is shown for representative Models 1, 2.1
and 5.1 (marked with filled diamonds) located around the centre, and
two of the corners of the box. In addition, Sequence 6 which crosses
the $\log g$--$T_{\rm eff}$ determination of \citet{fontaine08} is
also plotted in Fig.~\ref{fig:tracks}, and a model from this sequence
within their spectroscopic box analysed for comparison.

Table~\ref{tab:models} shows the models' $T_{\rm eff}$, $\log g$ and
envelope mass fractions, which, for the selected models (marked with
the initial of their plotting colour), are plotted in
Fig.~\ref{fig:models2}\footnote{All figures are plotted as a function
  of the fractional mass depth $\log q = \log (1 - M_r/M_T)$,
  which better samples the outer envelope of the star.}. Panel 1
shows that the cores of all models are devoid of hydrogen inside $\log q
\simeq -2.5$, where hydrogen shell burning begins. The models are also
devoid of helium in the core, until it starts to burn in a shell
around $\log q \simeq -1$. Panel 2 shows He$^3$ which is an
intermediate product of the proton-proton chain reaction (PP I)
produced in the hydrogen burning which creates He$^4$ nuclei. He$^3$
can also directly react with He$^4$ nuclei and, via intermediate
processes, produce two He$^4$ nuclei (PP II process). Panels 3 and 5
shows that the core of the models is composed of the residuals of the
helium burning: carbon, produced in the 3-$\alpha$ process, and
oxygen, as a result of the carbon capturing helium nuclei. N$^{14}$,
in Panel 4, is produced as a subproduct of the CNO cycle (C$^{12}$
recombining with protons, shown in the decrease of carbon at $\log q
\simeq -2$) which creates more C$^{12}$ and He$^4$. The last panel
display the mass fraction of all the remaining heavy elements.

   \begin{figure*}
   \begin{tabular}{ccc}
   \resizebox{0.45\linewidth}{!}{\includegraphics[angle=90]{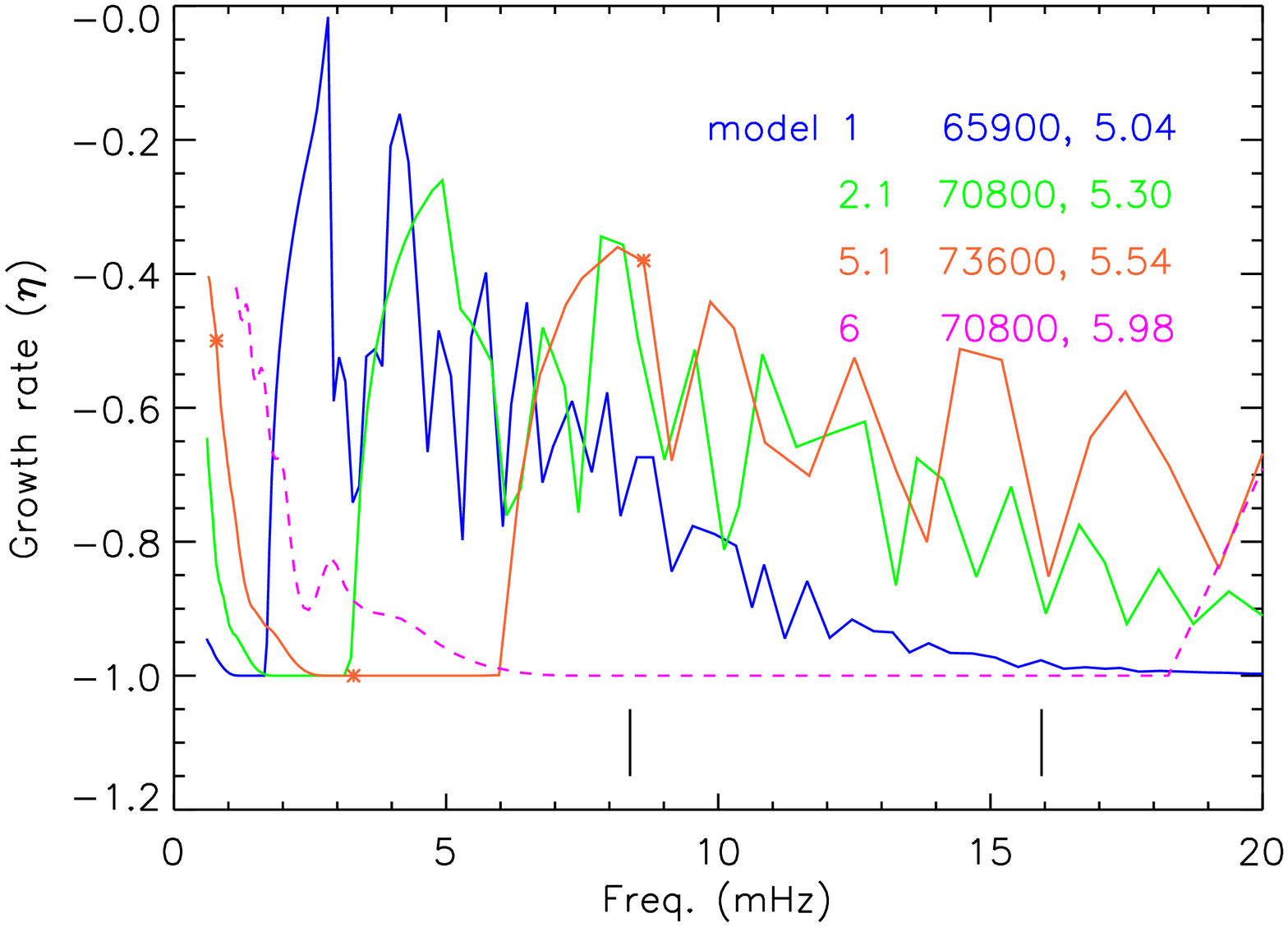}}
   &
   \resizebox{0.45\linewidth}{!}{\includegraphics[angle=90]{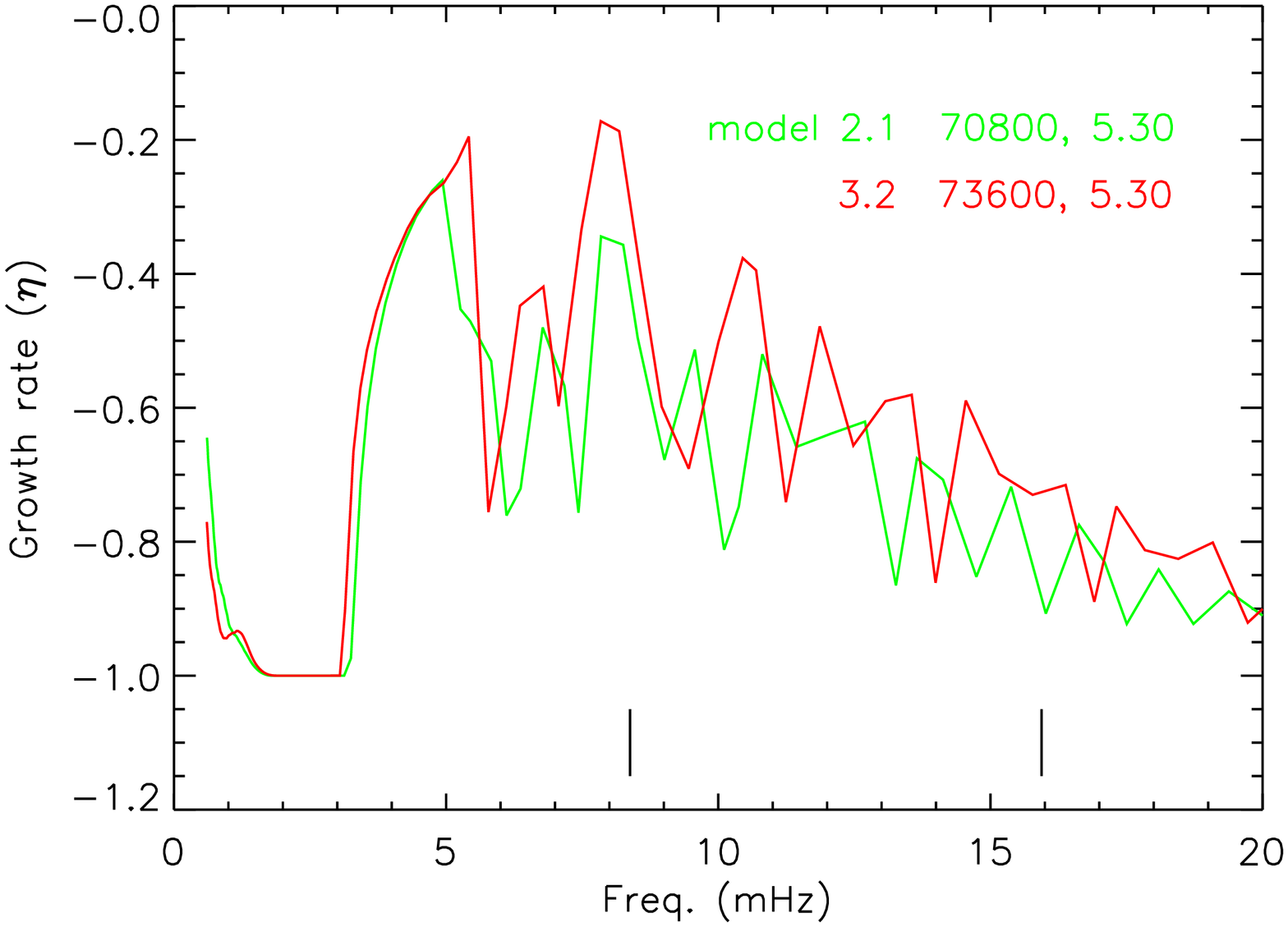}}
   \\
   \end{tabular}
   \caption{Left:Growth rates of representative models. Asterisks mark
     modes {\em g}76, {\em g}15 and {\em p}2 of Model 5.1,
     representative of its L, I and H regions, respectively. The two
     vertical dashes delimit the range of instability found for
     J\,16007+0748. Right: Growth rates of two models with same $\log
     g$ and different $T_{\rm eff}$. (See text for details).}
   \label{fig:grates}
   \end{figure*}
   \begin{figure*}
   \begin{tabular}{ccc}
   \resizebox{0.32\linewidth}{!}{\includegraphics[angle=90]{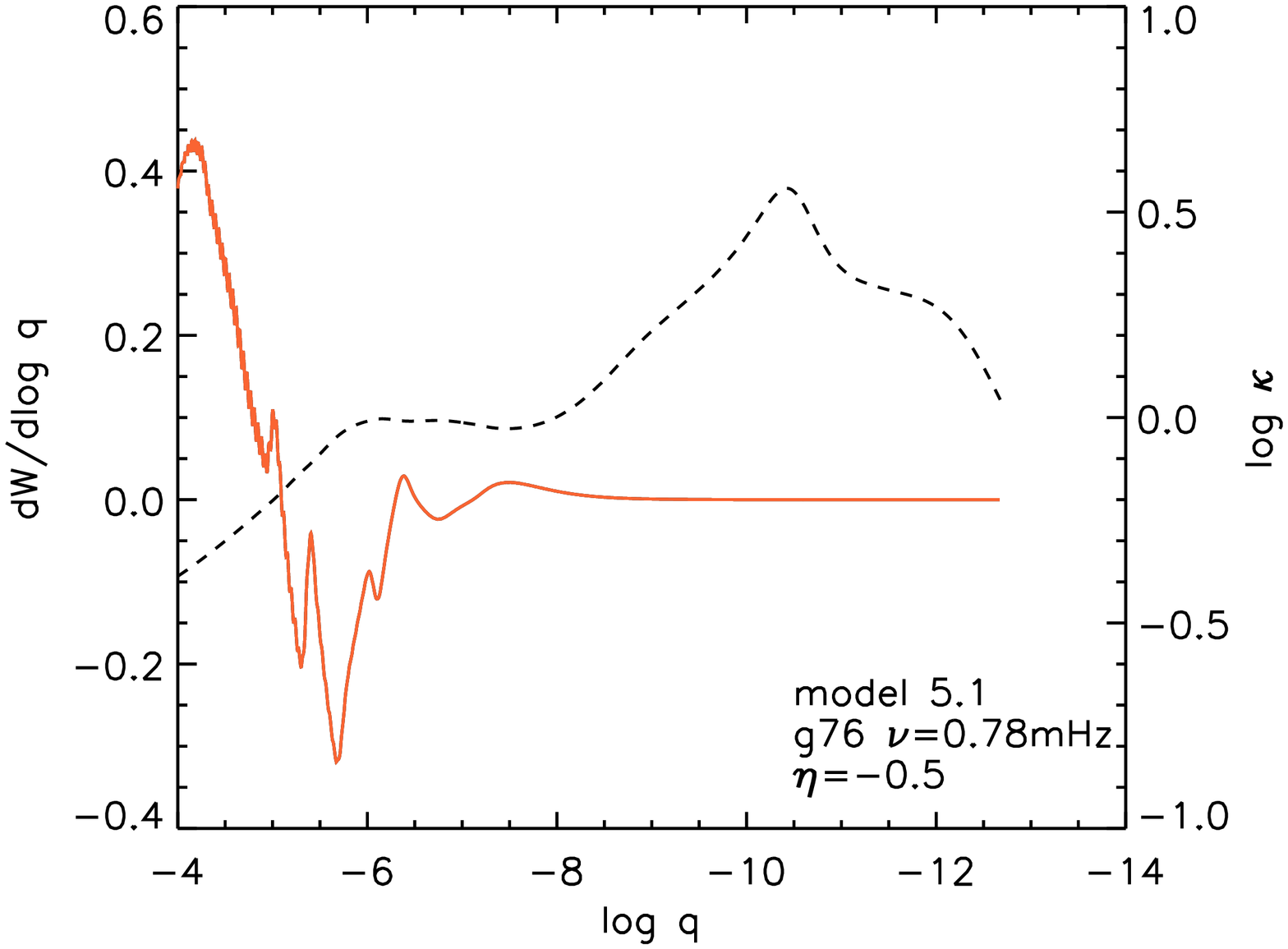}}
   \resizebox{0.32\linewidth}{!}{\includegraphics[angle=90]{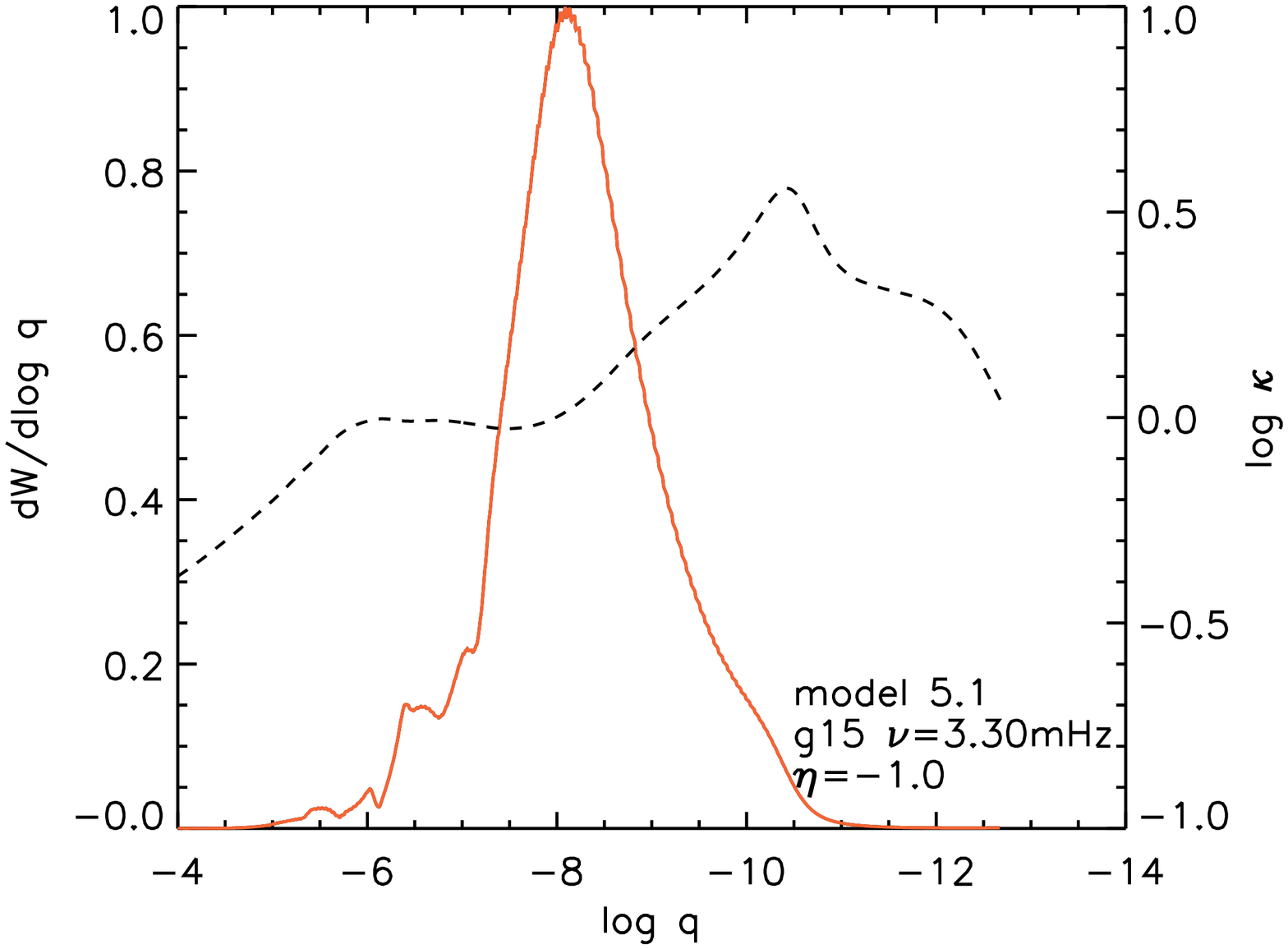}}
   &
   \resizebox{0.32\linewidth}{!}{\includegraphics[angle=90]{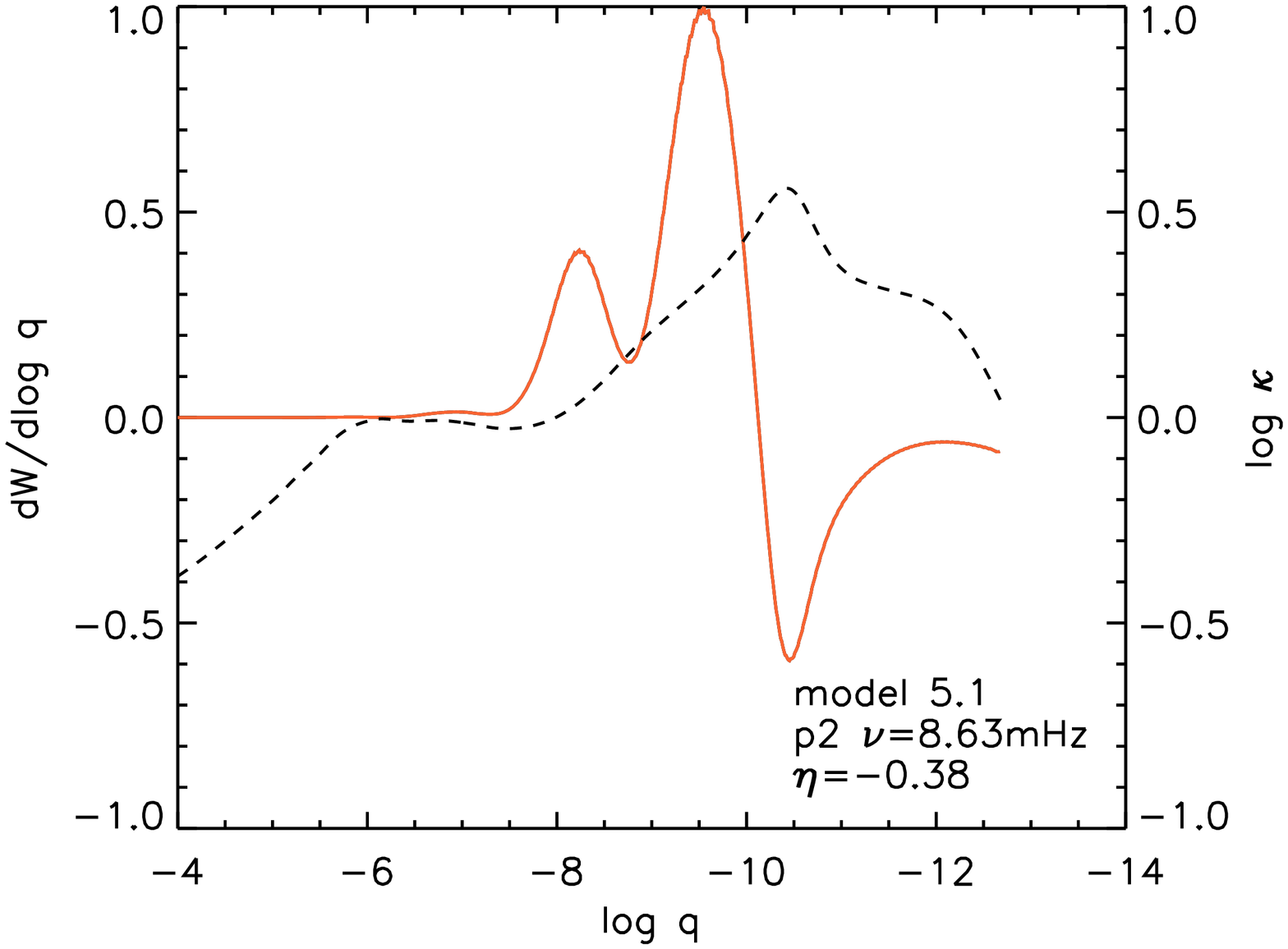}}
   \\
   \end{tabular}
   \caption{Left, Center \& Right: Energy and opacity for the
   quadrupole modes of Model 5.1 {\em g}76, {\em g}15 and {\em p}2,
   belonging to L, I and H zones, respectively (see text for
   details). All plots have been scaled to arbitrary units.}
   \label{fig:modes5.1}
   \end{figure*}

Some  representative physical  parameters of the selected models are
shown in Fig.~\ref{fig:phys22}. Panels 1 and 2 display the expected
behaviour of pressure and density, respectively. Panel 3 shows the
logarithm of the opacity: the largest opacity bump corresponds to the
heavy elements partial ionisation zone, and the secondary bump at
$\log q \simeq -6$ to the C/O partial ionisation zone. We note that
for Model 6 (magenta) the opacity is larger, although the derivative
of the opacity --implied in the $\kappa$ mechanism driving process--
is qualitatively similar. Panel 4 shows the nuclear energy generation
rate normalised to its maximum: the deepest peak corresponds to the
helium burning shell, while the shallowest is that of the hydrogen
burning shell. We note that for Model 6 the contribution of the helium
burning is very low compared to that of the hydrogen; and that
hydrogen burning takes place deeper in the rest of the
models. Panel 5 shows the luminosity normalised at the surface of the
star with the contributions of both burning shells; for all the
models, except Model 6, the luminosity achieves negative values at the
location of the He burning shell. This is due to the increase in the
helium burning rate produced by the extra-mixing of He, C, N and O in
the envelope. The energy so produced diffuses inwards to the
semi-degenerate material right below the burning shell, that expands
and cools producing the negative luminosities.

Finally, Panel 6 shows on one hand, the logarithm of
Brunt-V\"ais\"al\"a frequency (solid lines) which shows local maxima
at the zones with maximum composition gradients. The deepest and
largest maximum is due to the transition between the C/O core and the
He burning shell, while the secondary peak --stronger for Model 6, as
the transition is more abrupt (see Panel 1 in
Fig.~\ref{fig:models2})-- is due to the He/H transition. All the
energy transport in the star is done by radiation except at the
location of the heavy elements partial ionisation zone where $N^2 <
0$, meaning that the energy transport is more efficiently done by
convection. Panel 6 also shows the Lamb frequency (dashed lines) for $l=2$
modes.

The models so constructed were subject to a non adiabatic stability
analysis carried out with {\scriptsize GraCo} non adiabatic code of
oscillations \citep{moya08}. Modes were calculated from $l=0$ to $l=4$
in the 600-20\,000~$\mu$Hz frequency range, and the normalized growth
rate $\eta$ \citep{dziembowski93}, was used as an indicator of
stability. In our convention, positive (negative) values of the growth
rate mean the mode is unstable (stable). Fig.~\ref{fig:grates} (left)
shows the growth rate values of each of the selected models, all of
which were found stable in the analysed frequency range.

Each model presents three different regions according to the behaviour
of the growth rate.  Each of these regions cannot be delimited in
frequency, as they are model dependent, but it is useful to label them
generally as low (L), intermediate (I) and high frequency region (H)
to describe their general behaviour:
\begin{itemize}
\item The very low frequency region (L), with a slight tendency to
instability, comprises more modes (extending to higher frequencies)
and its driving is increased as the  model $\log g$ increases,
and it is only slightly temperature dependent.

\item The intermediate frequency region (I) is defined by growth rate
values of $\eta=-1$, indicating its high stability. This region also
shifts to higher frequencies and includes more modes as the $\log g$
of the models increases.

\item The high frequency region (H) corresponds to frequencies over
  the I region which have a certain tendency to instability. This
  region also shifts to higher frequencies and comprises a wider range
  in frequencies as the model $\log g$ increases. Besides, when the
  model $T_{\rm eff}$ increases at constant $\log g$, there is an
  overall increase in the tendency towards instability (see
  Fig.~\ref{fig:grates}, right). We also verified that, for models with
  equal metallicity, those with higher $T_{\rm eff}$ had
  somewhat higher growth rate values.
\end{itemize}

The described behaviour of the L, I and H regions is also found for
models within the same evolutionary sequence. Besides, as expected,
the frequency of the fundamental mode increases when the $\log g$, or
mean density of the model increases. Therefore, the high frequency
region is associated with low-radial order {\em g}-modes for the
models with lower $\log g$, and shifts into the {\em p}-mode region as
the model $\log g$ increases.

We found Model 5.1 to be most favoured for instability within the range of
pulsation frequencies observed for J16007+0748  (roughly 8 to 16~mHz),
this has the highest $\log g$ and opacity among our evolution sequences.

We present a stability analysis carried out for Model 5.1 as
representative of all the other extra-mixed models. We examine the
local energy exchange for modes belonging to L, I and H zones,
respectively, through the derivative of the work integral, $d W/d \log
q$, whose negative (positive) values indicate local contribution to
driving (damping) of the mode. Fig.~\ref{fig:modes5.1} shows the
energy balance for representative modes {\em g}76, {\em g}15 and {\em
  p}2, with the opacity plotted in the right axis for a better
understanding of the plot. Mode {\em g}76 (left) is representative of
the low frequency range, comprising from 0.6 to $\sim$2.5~mHz, where
modes achieve a certain destabilisation. This is caused by the
existence of a driving region at the location of the C/O partial
ionisation zone at $\log q \simeq -6$. Mode {\em g}15 (centre) belongs
to the intermediate frequency range, with cyclic frequencies between
about 2.5 to 6~mHz, where modes are highly stable. This is explained
as at these frequencies, the maximum energy interchange occurs in a
damping region. Mode {\em p}2 (right) is representative of the high
frequency range, from 6~mHz and above. The growth rate increases, and
so the tendency to driving, due to the existence of a driving region
at the location of the Z-bump at $\log q \simeq -10.5$ for this
model. The increase or decrease of the driving and damping regions
with frequency,  around the Z-bump, accounts for the oscillating
profile of the growth rate in this region.

By comparison, Model 6, belonging to Sequence 6 which passes through
the $\log g$--$T_{\rm eff}$ region which \citet{fontaine08} obtain,
presents only a  destabilisation for very low and very high
frequencies (in the range 20 to 50~mHz, out of the range of the plot),
due to the C/O and heavy elements partial ionisation zones,
respectively. The observed instability frequency region for
J\,16007+0748 is highly stable for this model, as any significant
energy dampes the modes.

\subsection{Discussion}

Our pulsational study has to be compared with the discovery of
the mechanism of oscillations of J16007+0748 by
\citet{fontaine08}. These authors achieve real driving of modes within
the instability range observed, using non-uniform metallicity models
as a result of the diffusive equilibrium between radiative levitation
and gravitational settling. The need for a non-uniform distribution of
iron with depth to achieve an accurate description of the instability
strip for sdBs had already been established by the same group
\citep{charpinet97}, and it has now been confirmed by
\citet{fontaine08} and \citet{charpinet09} to be a crucial ingredient
also to achieve real driving in sdO models. All of our models have
been built with uniform metallicity, and hence our inability to
achieve real driving of the modes.

However, from several tests carried out in our pulsation analysis we
still can conclude that for sdO models within our derived
spectroscopic box, those with $\log g$ values higher than 5.30 are
more likely to be driven in the observed frequency
range of J16007+0748. Besides, for models with the same $\log g$,
hotter models have a higher tendency to drive pulsations. Hence, the pulsation
analysis favours the higher effective temperature and higher $\log
g$ estimate within our spectroscopic box. Among the analysed models,
Model 5.1 with $T_{\rm eff}$=73\,600~K and $\log g=5.54$ was found the
most favoured for driving within the unstable frequency range observed for
J16007+0748.

\section{Conclusions} 

We have presented a spectroscopic analysis of J16007+0748, a binary
system containing the only pulsating sdO star known to date, and a
late type companion. A flux calibrated spectrum from the sixth data
release of the SDSS was used for the analysis. We used {\scriptsize
  TLUSTY} \citep{hubeny95} to compute a grid of non-LTE model stellar
atmospheres for the sdO comprising H and He, which were further used
with {\scriptsize SYNSPEC} \citep[][and references therein]{zboril96}
to compute a grid of non-LTE synthetic spectra. For the late-type
companion, synthetic spectra by \citet{martins05} were used, and a
null projected rotational velocity was assumed for both stars. A
genetic algorithm optimisation was used to estimate parameters for
both binary components given in Table 1.

We compared our results with atmospheric parameters derived by
\citet{fontaine08} and find that our $T_{\rm eff}$  estimate
is consistent, although with larger error limits; and our $\log g$
is lower than theirs. As for the cool companion type,
the difference in spectral type estimate may arise from the
fact that we have included an interstellar reddening factor,
which required a redder model for the companion.

A discussion of possible evolution scenarios for the sdO component
took our results into account. The derived atmospheric parameters
classify J16007+0748 as a helium-rich sdO, although showing $T_{\rm
  eff}$ and $\log g$ values above those canonically considered for
these objects \citep[see i.e.][]{stroeer07}. We considered J16007+0748
being a post-RGB object following an evolutionary track of
$0.4~M_{\odot}$ \citep{driebe98}, but it is not clear how the helium
enrichment would have occurred. We also considered the object to be a
post-AGB star following \citet{blocker95a, blocker95b} and
\citet{schonberner79, schonberner83} evolutionary tracks. For most of
the tracks the rapid evolution from the AGB to the pre-white dwarf
stage renders the observational detection unlikely. However, two
'born-again' AGB tracks from Bl\"ocker, with 0.5 and 0.6$~M_{\odot}$,
having longer evolutionary time, make the observational detection more
likely. The possibility of the system to be physically bounded, as it
seems to be implied in the derived radial velocities, is also
discussed. However, uncertainties in the derived absolute magnitudes
and stellar masses prevent us from a final conclusion.

We presented new and more extensive photometry gathered during
nineteen nights on the 1.9-m telescope at SAAO. The frequency analysis
recovers with certainty seven out of ten frequencies established in an
earlier paper. We conclude that the shorter data sets give better
results, possibly due to amplitude variations that may be caused by
the binary orbital motion, which needs further
investigation. Therefore, a spectroscopic determination of the orbital
period is very desirable.

Finally, a pulsation analysis of uniform-metallicity sdO models within
the atmospheric parameter range showed no modes were actually excited,
but useful conclusions were obtained.  We found models with $\log g >$
5.30 were more likely to be unstable at the frequency range observed
for J16007+0748. Besides, for models with the same $\log g$, higher
effective temperatures increase the overall driving rate.  Therefore,
among our models, the most favoured for instability within the
observed frequency range has atmospheric parameters $T_{\rm eff}$ =
73\,600 and log $g$ = 5.54. We remark that \citet{fontaine08} achieve
actual driving of {\em p}-modes within the instability range of
J16007+0748 with models including radiative levitation, which they
identify as key ingredient in driving the oscillations.

\section{Acknowledgments}

One of us (CR-L) is grateful for travel support provided by a Particle
Physics and Astronomy Research Council Visitors' Grant, held by the
University of Oxford. CR-L also acknowledges an \'Angeles Alvari\~no
contract under Xunta de Galicia and financial support from the
Spanish Ministerio de Ciencia y Tecnolog\'ia under project number
ESP2004-03855-C03-01. This paper is based partly on work supported
financially by the National Research Foundation of South Africa. We
thank the South African Astronomical Observatory for a generous
allocation of observing time.

Funding for the SDSS and SDSS-II has been provided by the Alfred P. Sloan Foundation, the Participating Institutions, the National Science Foundation, the U.S. Department of Energy, the National Aeronautics and Space Administration, the Japanese Monbukagakusho, the Max Planck Society, and the Higher Education Funding Council for England. The SDSS Web Site is http://www.sdss.org/.

The SDSS is managed by the Astrophysical Research Consortium for the Participating Institutions. The Participating Institutions are the American Museum of Natural History, Astrophysical Institute Potsdam, University of Basel, University of Cambridge, Case Western Reserve University, University of Chicago, Drexel University, Fermilab, the Institute for Advanced Study, the Japan Participation Group, Johns Hopkins University, the Joint Institute for Nuclear Astrophysics, the Kavli Institute for Particle Astrophysics and Cosmology, the Korean Scientist Group, the Chinese Academy of Sciences (LAMOST), Los Alamos National Laboratory, the Max-Planck-Institute for Astronomy (MPIA), the Max-Planck-Institute for Astrophysics (MPA), New Mexico State University, Ohio State University, University of Pittsburgh, University of Portsmouth, Princeton University, the United States Naval Observatory, and the University of Washington.

\bsp


\begin{thebibliography}{99}

\bibitem[\protect\citeauthoryear{Abbott}{1982}]{abbott82} Abbott D. C., 1982, \apj, 259, 282

\bibitem[\protect\citeauthoryear{Adelmann-McCarthy et al.}{2008}]{adelmann08} Adelmann-McCarthy K. et al., 2008, \apjs, 175, 297

\bibitem[\protect\citeauthoryear{Am{\^o}res \& L{\'e}pine}{2005}]{amores05} Am{\^o}res E. B., L{\'e}pine J. R. D., 2005, AJ, 130, 659

\bibitem[\protect\citeauthoryear{Avni}{1976}]{avni76} Avni Y., 1976, \apj, 210, 642


\bibitem[\protect\citeauthoryear{Bl\"ocker}{1995a}]{blocker95a} Bl\"ocker T., 1995a, A\&A, 297, 727

\bibitem[\protect\citeauthoryear{Bl\"ocker}{1995b}]{blocker95b} Bl\"ocker T., 1995b, A\&A, 299, 755

\bibitem[\protect\citeauthoryear{Bohlin \& Gilliland}{2004}]{bohlin04} Bohlin R. C., Gilliland R. L., 2004, AJ, 128, 3053

\bibitem[\protect\citeauthoryear{Charbonneau}{1995}]{charbonneau95} Charbonneau P., 1995, \apjs, 101, 309

\bibitem[\protect\citeauthoryear{Charpinet et al.}{1997}]{charpinet97} Charpinet S., Fontaine G., Brassard P., Chayer, P., Dorman B., 1997, \apj, 483, L123

\bibitem[\protect\citeauthoryear{Charpinet, Fontaine \& Brassard}{2009}]{charpinet09} Charpinet S., Fontaine G., Brassard P., 2009, A\&A, 493, 595

\bibitem[\protect\citeauthoryear{Deeming}{1968}]{deeming68} Deeming T. J., 1968, Vistas in Astr., 10, 125

\bibitem[\protect\citeauthoryear{Deeming}{1975}]{deeming75} Deeming T. J., 1975, Ap\&SS, 36, 137

\bibitem[\protect\citeauthoryear{D'Cruz et al.}{1996}]{dcruz96} D'Cruz N. L., Dorman B., Rood R. T., O'Connell R. W., 1996, \apj, 466, 371

\bibitem[\protect\citeauthoryear{Driebe et al.}{1998}]{driebe98} Driebe T., Sch{\"o}nberner D., Bl{\"o}cker T., Herwig F., 1998, A\&A, 339, 123

\bibitem[\protect\citeauthoryear{Driebe et al.}{1999}]{driebe99} Driebe T., Sch{\"o}nberner D., Bl{\"o}cker T., Herwig F., 1999, A\&A, 350, 89

\bibitem[\protect\citeauthoryear{Dziembowski et al.}{1993}]{dziembowski93} Dziembowski W. A., Moskalik, P., Pamyatnykh A. A., 1993, \mnras, 265, 588

\bibitem[\protect\citeauthoryear{Fontaine et al.}{2008}]{fontaine08} Fontaine G., Brassard P., Green E. M., Chayer P., Charpinet S., Andersen M., Portouw J., 2008, A\&A, 486, L39

\bibitem[\protect\citeauthoryear{Howarth}{1983}]{howarth83} Howarth I. D., 1983, MNRAS, 203, 301

\bibitem[\protect\citeauthoryear{Hu et al.}{2008}]{hu08} Hu H., Dupret M.-A., Aerts C., Nelemans G., Kawaler S. D., Miglio A., Montalban J., Scuflaire R., 2008, A\&A, 490, 243

\bibitem[\protect\citeauthoryear{Hubeny \& Lanz}{1995}]{hubeny95} Hubeny I., Lanz T., 1995, ApJ, 439, 875

\bibitem[\protect\citeauthoryear{Hubeny, Hummer \& Lanz}{1994}]{hubeny94} Hubeny I., Hummer D. G. \& Lanz T., 1994, A\&A, 282, 151


\bibitem[\protect\citeauthoryear{Kilkenny et al.}{1998}]{kilkenny98} Kilkenny D., O'Donoghue D., Koen C., Lynas-Gray A.E., van Wyk F., 1998, MNRAS, 296, 329

\bibitem[\protect\citeauthoryear{Kilkenny et al.}{2003}]{kilkenny03} Kilkenny D. et al., 2003, MNRAS, 345, 834

\bibitem[\protect\citeauthoryear{Kurtz}{1985}]{kurtz85} Kurtz D. W., 1985, MNRAS, 213, 773

\bibitem[\protect\citeauthoryear{Lawlor \& MacDonald}{2006}]{lawlor06} Lawlor T. M., MacDonald J., 2006, \mnras, 371, 263

\bibitem[\protect\citeauthoryear{Martins et al.}{2005}]{martins05} Martins L. P., Delgado R. M. G., Leitherer C., Cervi{\~n}o M., Hauschildt P., 2005, MNRAS, 385, 49

\bibitem[\protect\citeauthoryear{Moehler et al.}{1990}]{moehler90} Moehler S., Richtler T., de Boer K., Dettmar R. J., Heber U., 1990, A\&AS, 86, 53

\bibitem[\protect\citeauthoryear{Mortimore \& Lynas-Gray}{2006}]{mortimore06} Mortimore A. N., Lynas-Gray A.E., 2006, Baltic Astr., 15, 207

\bibitem[\protect\citeauthoryear{Moya \& Garrido}{2008}]{moya08} Moya A., Garrido R., 2008, \apss, 316, 129

\bibitem[\protect\citeauthoryear{O'Donoghue}{1995}]{odonoghue95} O'Donoghue D., 1995, Baltic Astr., 4, 519

\bibitem[\protect\citeauthoryear{O'Donoghue et al.}{1997}]{odonoghue97} O'Donoghue D., Lynas-Gray A. E., Kilkenny D., Stobie R.S., Koen C, 1997, MNRAS, 285, 657

\bibitem[\protect\citeauthoryear{Rodr\'iguez-L\'opez et al.}{2008}]{rod08} Rodr\'iguez-L\'opez C., Garrido R., Moya A., MacDonald J., Ulla A., 2008, in Heber U., Jeffery C. S. \& Napiwotzki R., eds., ASP Conf. Ser. Vol. 392, Hot Subdwarf Stars and Related Objects. Astron. Soc. Pac., San Francisco, p. 363.

\bibitem[\protect\citeauthoryear{Saffer et al.}{1994}]{saffer94} Saffer R.A., Bergeron P., Koester D., Liebert J., 1994, ApJ, 432, 351

\bibitem[\protect\citeauthoryear{Seaton}{1979}]{seaton79} Seaton M. J., 1979, MNRAS, 187, 73p

\bibitem[\protect\citeauthoryear{Schechter, Mateo \& Saha}{Schechter et al.}{1993}]{schechter93} Schechter P. L., Mateo M., Saha A., 1993, PASP, 105, 1342

\bibitem[\protect\citeauthoryear{Sch\"onberner}{1979}]{schonberner79} Sch{\"o}nberner D., 1979, A\&A, 79, 108

\bibitem[\protect\citeauthoryear{Sch\"onberner}{1983}]{schonberner83} Sch{\"o}nberner D., 1983, ApJ, 272, 708

\bibitem[\protect\citeauthoryear{Sch\"oning}{1994}]{schoning94} Sch{\"o}ning T., 1994, A\&A, 282, 994

\bibitem[\protect\citeauthoryear{Sch\"oning \& Butler}{1989}]{schoning89} Sch{\"o}ning T., Butler K., 1989, A\&AS, 78, 51

\bibitem[\protect\citeauthoryear{Simon \& Sturm}{1994}]{simon94} Simon K. P., Sturm E., 1994, A\&A, 281, 286

\bibitem[\protect\citeauthoryear{Smith et al.}{2002}]{smith02} Smith J. A. et al., 2002, AJ, 123, 2121 

\bibitem[\protect\citeauthoryear{Stroeer et al.}{2007}]{stroeer07} Stroeer A., Heber U., Lisker T., Napiwotzki R., Dreizler S., Christlieb N., Reimers D., 2007, A\&A, 462, 269

\bibitem[\protect\citeauthoryear{Vidal, Cooper \& Smith}{Vidal et al.}{1970}]{vidal70} Vidal C. R., Cooper, J., Smith E. W., 1970, JQRST, 10, 1011

\bibitem[\protect\citeauthoryear{Wald \& Wolfowitz}{1940}]{wald40} Wald A., Wolfowitz J., 1940, Ann. Math. Stat., 11, 147 

\bibitem[\protect\citeauthoryear{Woudt et al.}{2006}]{woudt06} Woudt P. A. et al., 2006, MNRAS, 371, 1497

\bibitem[\protect\citeauthoryear{Zboril}{1996}]{zboril96} Zboril M., 1996, in``Model Atmospheres and Spectrum Synthesis'', eds. Adelman S. J., Kupka F., Weiss W. W., ASP Conf. Ser. 108, p.193

\end{thebibliography}
\end{document}